\magnification=1200
\parskip=10pt
\parindent=15pt
\baselineskip=13pt \pageno=0
\footline={\ifnum \pageno <1 \else\hss\folio\hss\fi}
\input amssym.def
\input amssym
\def\ssquare{\vrule width.6em height.5em depth.1em \relax}
\def\qed{\ifhmode\unskip\fi\hfil\ssquare}
\line{\hfil{RFSC 04-01}}
\line{\hfil{Revised~~~~~~}} \vskip 1in \centerline{\bf A PROOF OF THE
ODD PERFECT NUMBER CONJECTURE} \vskip .5in \centerline{Simon Davis}
\vskip .5in {\bf Abstract.}  It is sufficient to prove that there is an excess of prime 
factors in the product of repunits with odd prime bases defined by the sum of divisors 
of the integer $N=(4k+1)^{4m+1}\prod_{i=1}^\ell ~ q_i^{2\alpha_i}$ to establish that
there do not exist any odd integers with equality between $\sigma(N)$ and 2N.
The existence of distinct prime divisors in the repunits in $\sigma(N)$ follows
from a theorem on the primitive divisors of the Lucas sequences 
$U_{2\alpha_i+1}(q_i+1,q_i)$ and $U_{2\alpha_j+1}(q_j+1,q_j)$ with
$q_i,q_j,2\alpha_i+1,2\alpha_j+1$ being odd primes. The occurrence of new prime
divisors in each quotient ${{(4k+1)^{4m+2}-1}\over {4k}}$,
${{q_i^{2\alpha_i+1}-1}\over {q_i-1}}, i=1,...,\ell$ also implies that the
square root of the product of $2(4k+1)$ and the sequence of repunits will not
be rational unless the primes are matched. Although a finite set of solutions to the 
rationality condition for the existence of odd perfect numbers is obtained, it is 
verified that they all satisfy ${{\sigma(N)}\over N}\ne 2$ because the repunits in 
the product representing $\sigma(N)$ introduce new prime divisors.  Minimization of 
the number of prime divisors in $\sigma(N)$ leads to an infinite set of repunits of 
increasing mangitude or prime equations with no integer solutions. It is proven then 
that there exist no odd perfect numbers.

\vskip .8in \noindent {\bf MSC Class: 11D61, 11K65}
\vfill\eject \noindent {\bf 1. Introduction} \vskip 10pt

While even perfect numbers were known to be given by
$2^{p-1}(2^p-1)$, for $2^p-1$ prime, the universality
of this result led to the the problem of
characterizing any other possible types of perfect
numbers.  It was suggested initially
by Descartes that it was not likely that odd integers could be
perfect numbers [12].  After the work of de Bessy [3],
Euler proved that the condition ${{\sigma(N)}\over N}=2$, where
 $\sigma(N)
=\sum_{{d\vert N}\atop {d~integer}} d$ is the sum-of-divisors
function, restricted odd integers to have the form
$(4k+1)^{4m+1}q_1^{2\alpha_1}...q_\ell^{2\alpha_\ell}$, with
$4k+1,~q_1,...,q_\ell$ prime [13], and further, that there might
exist no set of prime bases such that the perfect number
condition was satisfied.

Investigations of the equation for the sum of the reciprocals of the
divisors of $N$ has led to lower bounds on the number of
distinct prime divisors.  This number has increased from four to nine
[17][27][33] while a minimum of $75$ total prime factors [19] has been
established.   When $3\not\mid N$, it was shown that there would be a
 minimum
of twelve different prime divisors [18][24].  It was demonstrated also
 that, if
$3,~5,~7\not\mid N$, greater than $26$ distinct prime factors
would be required [7][28].  In decreasing order, the three largest
prime divisors were bounded below by $10^8+7$ [15], $10^4+7$ [20] and
 $10^2+1$
[21] respectively, while the least prime divisor had to be less than
${{2n+6}\over 3}$ for $n$ different prime factors [16] and
 $exp(4.97401\times 10^{10})$
[34].  Moreover, either one of the prime powers, $(4k+1)^{4m+1}$ or
$q_i^{2\alpha_i}$ for some index $i$, was found to be larger than
 $10^{20}$ [10].
Through the algorithms defined by the sum over reciprocals of the
 divisors, it has
been demonstrated that odd perfect numbers had to be greater than
 $10^{300}$ [6].

One of the possible methods of proof of the odd perfect number
conjecture is based on the harmonic mean $H(N)={{\tau(N)}
\over {\sum_{d\vert N}{1\over d}}}$,
where $\tau(N)$ is the number of integer divisors of $N$.  It has
been conjectured that $H(N)$ is not integer when $n$ is odd [30].  This
statement also would imply the nonexistence of odd perfect numbers as
the perfect number condition is $\sum_{d\vert N} {1\over d}=2$ and
$\tau(N)$ must be even, since $N$ is not a perfect square.  The
use of the harmonic mean leads again to the study of the sum of
the reciprocals of the divisors, and the values of this sum have
only been approximated.  For example, it has been found that there
are odd integer with five different prime factors such that
$\left\vert {{\sigma(N)}\over N}-2\right\vert < 10^{-12}$ [23].

The uniqueness of the prime decomposition of an integer allows for the
comparison between its magnitude and the sum of the divisors. Since
the sum of the divisors of an odd integer $N=(4k+1)^{4m+1}~
\prod_{i=1}^\ell~q_i^{2\alpha_i}$ by the product
${{(4k+1)^{4m+2}-1}\over {4k}}~
\prod_{i=1}^\ell~{{q_i^{2\alpha_i+1}-1}\over {q_i-1}}$, it is
sufficient to determine the properties of the prime factors of the
repunits to establish that there exist no odd perfect numbers. The
irrationality of ${\sqrt{2(4k+1)}}\bigg[\sigma((4k+1)^{4m+1})\cdot
\prod_{i=1}^\ell~$\hfil\break $\sigma(q_i^{2\alpha_i})\bigg]^{1\over
2}$ would imply that $\sigma(N)$ cannot equal $2N$.   It has been
proven for a large class of primes $\{4k+1;q_i\}$ and exponents
$\{4m+1;2\alpha_i\}$ that the rationality condition is not
satisfied. The irrationality of the square root of the product of
$2(4k+1)$ and the sequence of repunits is not valid for all sets of
primes and exponents, however, and it is verified in $\S 4$ that the
rationality condition holds for twelve odd integers.  The
factorizations of these integers have the property that the repunits
have prime divisors which form interlocking rings, whereas, in
general, the sequence of prime factors does not close. The presence
of a sequence of primes of increasing magnitude prevents any finite
odd integer from being a perfect number. To prove that
${{\sigma(N)}\over N}\ne 2$, it is necessary also to obtain a lower
bound for the number of prime divisors in $\sigma(N)$.  Since the
number of distinct prime factors of ${{q_i^n-1}\over {q_i-1}}$ is
minimized in the class of repunits with exponents containing the
prime divisor $p$ when $n=p$, the exponent is presumed to
be prime throughout the discussion.  It is demonstrated in Theorem 1
that any pair of repunits, with prime bases and exponents satisfying
a given inequality, do not have identical sets of prime divisors.
Then, either $\sigma(N)$ has an excess of prime divisors or
constraints must be imposed on $\{4k+1;q_i\}$ and
$\{4m+1;2\alpha_i\}$ which have no integer solution.  The
non-existence of odd perfect numbers also follows from the sequence
of prime factors of increasing magnitude in the factorization of
$\sigma(N)$, when one of three specified relations is satisfied, and
constraints on the basis and exponents otherwise. \vskip 10pt
\noindent {\bf 2. The Existence of Different Prime Divisors in the
Repunit Factors of the} \hfil\break \noindent \phantom{.....}{\bf
Sum of Divisors} \vskip 10pt

To prove the nonexistence of odd perfect numbers, it shall
be demonstrated the product of repunits in the expression for
 $\sigma(N)$
contains an excess of prime divisors of $\sigma(N)$,
such that the perfect number condition $\sigma(N)=2N$
cannot be satisfied.
The existence of distinct prime divisors in the quotients
${{(4k+1)^{4m+2}-1}\over {4k}}$ and ${{q_i^{2\alpha_i+1}-1}\over
{q_i-1}},~i=1,..., \ell$ follows from a theorem on the prime
factors of the quotients ${{q_i^{2\alpha_i+1}-1}\over {q_i-1}}$
where $q_i$ and $2\alpha_i+1$ are odd primes.  A restriction
to the least possible number of distinct prime factors in
a product ${{q_i^{2\alpha_i+1}-1}\over {q_i-1}}\cdot
{{q_j^{2\alpha_j+1}-1}\over {q_j-1}}$ yields the three
equations, where the complement has been considered previously [11].
It is shown that the third constraint cannot be satisfied if
 $\alpha_i=\alpha_j$,
and a new prime divisor occurs in one of the repunits if the other two
constraints equations or $\alpha_i\ne \alpha_j$ in the third condition.

If the inequality
$${{q_i^{2\alpha_i+1}-1}\over {q_i-1}}\ne {{q_j^{2\alpha_j+1}-1}\over
 {q_j-1}}
\eqno(2.1)
$$
holds, then either the sets of primitive
divisors of ${{q_i^{2\alpha_i+1}-1}\over {q_i-1}}$ and
${{q_j^{2\alpha_j+1}-1}\over {q_j-1}}$ are not identical or the
exponents of the prime power divisors are different, given that
${{q_i^{2\alpha_i+1}-1}\over {q_i-1}}\ne {{q_j^{2\alpha_j+1}-1}\over
{q_j-1}}$. Let $q_i>q_j$.  Since
$$\eqalign{
{{q_i^{2\alpha_i+1}-1}\over {q_j^{2\alpha_j+1}-1}}&=
{{q_i^{2\alpha_i+1}-1} \over
{q_i^{2\alpha_j+1}-1}}{{q_i^{2\alpha_j+1}-1}\over
{q_j^{2\alpha_j+1}-1}} \cr &= {{q_i^{2\alpha_i+1}-1}\over
{q_i^{2\alpha_j+1}-1}} \bigg[\left({{q_i}\over
{q_j}}\right)^{2\alpha_j+1}
\bigg(1+{{q_i^{2\alpha_i+1}-q_j^{2\alpha_j+1}-1}\over
{q_i^{2\alpha_j+1} q_j^{2\alpha_j+1}}}+{1\over
{q_j^{4\alpha_j+2}}}\cr &~~~~~~~~~~~~~~~~~~~~~~~~~~~~~~~~~~~~~
~~~~~~~~~~~~~~~~~~~~~-{1\over
{q_i^{2\alpha_j+1}q_j^{4\alpha_j+2}}}+...\bigg)\bigg] \cr}
\eqno(2.2)
$$
this equals
$$\eqalign{
&{{q_i^{2\alpha_i+1}-1}\over {q_i^{2\alpha_j+1}-1}}
\left[\left({{q_i}\over {q_j}}\right)^{2\alpha_j+1}+\epsilon\right]
\cr &\epsilon\simeq \left({{q_i}\over
{q_j}}\right)^{2\alpha_j+1}\left(
{{q_i^{2\alpha_j+1}-q_j^{2\alpha_j+1}-1}\over
{q_i^{2\alpha_j+1}q_j^{2\alpha_j+1}}}+{1\over
{q_j^{4\alpha_j+2}}}+...\right) \cr} \eqno(2.3)
$$
When $q_j < q_i < \left[q_j^{2\alpha_j+1}\cdot
(q_j^{2\alpha_j+1}-1)\right]^{{{2\alpha_j+1}\over
{2\alpha_i+1}}\left[1+{{(2\alpha_j+1)^2} \over
{2\alpha_i+1}}\right]^{-1}}$, $\epsilon \cdot
{{q_i^{2\alpha_i+1}-1}\over {q_i^{2\alpha_j+1}-1}}< 1$.  The exact
value of the remainder term is
$$\epsilon\cdot {{q_i^{2\alpha_i+1}-1}\over {q_i^{2\alpha_j+1}-1}}
={{q_i^{2\alpha_i+1}-1}\over {q_i^{2\alpha_j+1}-1}}
{{q_i^{2\alpha_i+1}-q_j^{2\alpha_j+1}}\over
{q_j^{2\alpha_j+1}(q_j^{2\alpha_j+1}-1)}} \eqno(2.4)
$$
which yields the alternative conditions $q_i< 2^{-{1\over
{2\alpha_i+1-2(2\alpha_j+1)}}} q_j^{{2\alpha_j+1}\over
{2\alpha_i+1-2(2\alpha_j+1)}}$ and \hfil\break $q_i<{1\over
2}q_j^{3{{2\alpha_i+1}\over {2\alpha_j+1}}}$ for the remainder term
to be less than 1.  Since the entire quotient is
${{q_i^{2\alpha_i+1}-1}\over {q_i^{2\alpha_j+1}-1}}
\bigg[{{q_j^{4\alpha_j+2}(q_j^{2\alpha_j+1}-1)+q_i^{2\alpha_j+1}-q_j^{2\alpha_j+1}}
\over
{q_i^{2\alpha_j+1}q_j^{2\alpha_j+1}(q_j^{2\alpha_j+1}-1)}}\bigg]$,
it can be an integer or a fraction with the denominator introducting
no new divisors other than factors of $q_i^{2\alpha_i-1}$ only if
$q_j^{2\alpha_j+1}-q_j^{2\alpha_j+1}$ contains all of the divisors
of $q_j^{2\alpha_j+1}-1$, presuming that $q_j^{2\alpha_j+1}$ is
cancelled.  Similarly, if $q_j < q_i
\left[q_j^{2\alpha_j+1}(q_j^{2\alpha_j+1}-1)\right]^{{{2\alpha_j+1}\over
{2\alpha_i+1}}\left[1+{{(2\alpha_j+1)^2}\over
{2\alpha_i+1}}\right]^{-1}}$, the remainder term must be fractional
and ${{q_j^{2\alpha_j+1}-1}\over {q_i-1}}$ must have a higher power
of one of the prime divisors or a distinct prime divisor,
unless $q_i^{2\alpha_j+1}-1$ has a factor which
is different from the divisors of $q_j^{2\alpha_j+1}-1$, which
typically occurs if the odd primes $q_i,~q_j$ are sufficiently large
and $\alpha_i,~\alpha_j\ge 1$, as ${{q_i^{2\alpha_j+1}-1}\over
{q_j^{2\alpha_j+1}-1}}\simeq \left({{q_i}\over
{q_j}}\right)^{2\alpha_j+1}$.
\hfil\break\hfil\break
{\bf Example 1.}  Consider the prime divisors of pairs of repunits
and the integrality of the quotients ${{q_i^n-1}\over {q_j^n-1}}$ when
$q_i,~q_j$ and $n$ are odd primes.  The first example of an integer
 ratio
for odd prime bases and exponents is
${{q_i^{2\alpha_i+1}-1}\over {q_j^{2\alpha_j+1}-1}}={{29^3-1}\over
 {3^3-1}}=938$,
but the quotient ${{29^3-1}\over {29-1}}=13\cdot 67$ introduces a
new prime divisor.  For $q_i,~q_j\gg 1$, the ratio
 ${{q_i^{2\alpha_i+1}-1}
\over {q_j^{2\alpha_i+1}-1}}\to \left({{q_i}\over
 {q_j}}\right)^{2\alpha_i+1}$,
which cannot be integer for odd primes $q_i,~q_j$, implying that one
of the repunits, with $\alpha_i=\alpha_j$, has a new prime divisor
or $q_j^{2\alpha_i+1}-1$ contains a higher power of one of the prime
 factors.
The latter possibility is excluded because, the numerator and
 denominator
are raised to the same power in the limit of large $q_i,~q_j$.

\line{\hfil{$\square$}}

\hfil\break \hfil\break
{\bf Theorem 1.} The pair of integers
${{q_i^{2\alpha_i+1}-1}\over {q_i-1}}$ and
${{q_j^{2\alpha_j+1}-1}\over {q_j-1}}$, $q_i> q_j$ do not have
identical sets of prime divisors if the bases and exponents are odd
primes. \hfil\break \hfil\break {\bf Proof.} There are three known
positive integer solutions to the exponential Diophantine equation
${{x^m-1}\over {y^n-1}}$, $m,n>1$, $2^3-1={{6^2-1}\over {6-1}}$,
$31=2^5-1={{5^3-1}\over {5-1}}$ and $8191=2^{13}-1={{90^3-1}\over
{90-1}}$ [14][29][31] and no equalities of this type have been
established yet for odd prime bases and exponents.  It will be
assumed initially that
${{q_i^{2\alpha_i+1}-1}\over {q_i-1}}\ne {{q_j^{2\alpha_j+1}-1}\over
 {q_j-1}}$,
until this inequality is an evident consequence of the proof.

Let $q_i-1=p_1^{h_{i1}}...p_s^{h_{is}}$ and
$q_j-1=p_1^{h_{j1}}...p_s^{h_{js}}$. Then
$$\eqalign{
{{(p_1^{h_{i1}}...p_s^{h_{is}}+1)^{2\alpha_i+1}-1}\over
{p_1^{h_{1s}}...p_s^{h_{is}}}}
&=2\alpha_i+1+\alpha_i(2\alpha_i+1)(p_1^{h_{i1}}...p_s^{h_{is}})+...
+(p_1^{h_{i1}}...p_s^{h_{is}})^{2\alpha_i} \cr
{{(p_1^{h_{j1}}...p_s^{h_{js}}+1)^{2\alpha_j+1}-1}\over
{p_1^{h_{1s}}...p_s^{h_{js}}}} &=
2\alpha_j+1+\alpha_j(2\alpha_j+1)(p_1^{h_{j1}}...p_s^{h_{is}})
+...+(p_1^{h_{j1}}...p_s^{h_{js}})^{2\alpha_j} \cr} \eqno(2.5)
$$
Let $P$ be a primitive divisor of the $q_i^{2\alpha_i+1}-1$.
Consider the congruence
$c_0+c_1x+...+c_{2\alpha_i}x^{2\alpha_i}\equiv 0~(mod~P), c_t>0,$
\hfil\break $t=0,1,2,...,2\alpha_j$. There is a unique solution
$x_0$ to the congruence relation $f(x)\equiv c_0+c_1x+...+c_n
x^n\equiv 0~(mod~P)$ within $[x_0-\epsilon,x_0+\epsilon]$ [24],
where
$$\epsilon f^\prime(x_0)+{{\epsilon^2}\over
 {2!}}f^{\prime\prime}(x_0)+... ~<~ P
\eqno(2.6)
$$
Setting $c_t=\left({{2\alpha_i+1}\atop {t+1}}\right)$,
$$f^\prime(q_i-1)~=~\left({{2\alpha_i+1}\atop
 2}\right)+2\left({{2\alpha_i+1}\atop
3}\right)(q_i-1)+...+2\alpha_i(q_i-1)^{2\alpha_i-1} \eqno(2.7)
$$
the constraint (2.6) implies a bound on $\epsilon$ of
$\left(1-{2\over {2\alpha_i+1}}\right){{q_i-1}\over {2\alpha_i}}$.
Conseqently, if
$$\vert q_i- q_j\vert < \left(1-{2\over
 {2\alpha_i+1}}\right){{q_i-1}\over
{2\alpha_i}} \eqno(2.8)
$$
the prime $P$ is not a common factor of both repunits.  The number
of repunits with primitive divisor $P$ will be bounded by the
number of integer solutions to the congruence relation.

For $P$ to be a divisor of $q_j^{2\alpha_j+1}-1$,
$p_1^{h_{j1}}p_2^{h_{j2}}... p_s^{h_{js}}$ also would have to
satisfy the congruence relation when $\alpha_i=\alpha_j$. If
$p_1^{h_{i1}}p_2^{h_{i2}}...p_s^{h_{is}}=p_1^{h_{j1}}p_2^{h_{j2}}...
p_s^{h_{js}}+{\it n}P$, then
$p_1^{h_{j1}}p_2^{h_{j2}}...p_s^{h_{js}}$ is another solution to the
congruence. Therefore $gcd(q_i-1,q_j-1)=$ \hfil\break
$p_1^{min(h_{i1},h_{j1})}...p_s^{min(h_{is},h_{js})} \vert {\it
n}P$, and, as $P$ cannot equal any of the prime divisors $p_{ik}$,
$gcd(q_i-1,q_j-1)=$ \hfil\break
$p_1^{min(h_{i1},h_{j1})}...p_s^{min(h_{is},h_{js})}\vert {\it n}$.
Then, $q_i-1=q_j-1+{\it k}gcd(q_i-1,q_j-1)P$ for some integer ${\it
k}$, and $P\vert {{q_i-q_j}\over {gcd(q_i-1,q_j-1)}}$.

However, if $P$ is a prime divisor of ${{q_i^{2\alpha_i+1}-1}\over
{q_i-1}}$ and ${{q_j^{2\alpha_j+1}-1}\over {q_j-1}}$, and
$\alpha_i=\alpha_j$,
$$P\vert
 (q_i-q_j)(q_i^{2\alpha_i}+(2\alpha_i+1)q_i^{2\alpha_i-1}q_j+...+
(2\alpha_i+1)q_iq_j^{2\alpha_i-1}+q_j^{2\alpha_i}) \eqno(2.9)
$$
Since
$gcd(q_i-q_j,q_i^{2\alpha_i}+(2\alpha_i+1)q_i^{2\alpha_i-1}q_j+...+
(2\alpha_i+1)q_iq_j^{2\alpha_i-1}+q_j^{2\alpha_i})=1$, either
$P\vert q_i-q_j$ or \hfil\break $P\vert
q_i^{2\alpha_i}+(2\alpha_i+1)q_i^{2\alpha_i-1}q_j+...+
(2\alpha_i+1)q_iq_j^{2\alpha_i-1}+q_j^{2\alpha_i}$.  When $P\vert
q_i-q_j$, it may satisfy the divisibility condition for the
congruence relation.  However, there also would be a primitive
divisor which is a factor of
$q_i^{2\alpha_i}+q_i^{2\alpha_i-1}q_j+...+q_j^{2\alpha_i-1}q_i+q_j^{2\alpha_i}$,
and not $q_i-q_j$.  Then $p_1^{h_{i1}}p_2^{h_{i2}}...p_s^{h_{is}}
< p_1^{h_{j1}}p_2^{h_{j2}}...p_s^{h_{js}}+P$.  If there is only
one solution to the congruence
$2\alpha_i+1+\alpha_i(2\alpha_i+1)x+...+x^{2\alpha_i}\equiv 0
~(mod~P)$ less than $P$, a contradiction is obtained. When one of
the congruences is not satisfied for at least one primitive
divisor $P$, this prime is not a common factor of both repunits.

If $\alpha_i\ne \alpha_j$,
$$\eqalign{
q_i^{2\alpha_i+1}-1&=\prod_{k=0}^{2\alpha_i}~(q_i-\omega_{2\alpha_i+1}^k)
\cr q_j^{2\alpha_j+1}-1&=\prod_{k^\prime=0}^{2\alpha_j}~
(q_j-\omega_{2\alpha_j+1}^{k^\prime}) \cr \omega_n&=e^{{2\pi i}\over
n} \cr} \eqno(2.10)
$$
Quotients of products of linear factors in $q_i^{2\alpha_i+1}-1$
and $q_j^{2\alpha_j+1}-1$ generally will not involve cancellation
of integer divisors because products of powers of the different
units of unity cannot be equal unless the exponents are a
multiples of the primes $2\alpha_i+1$ or $2\alpha_j+1$.  The
sumsets which give rise to equal exponents can be enumerated by
determining the integers $s_1,...,s_t$ such that the sums are
either congruent to 0 modulo $2\alpha_i+1$ and $2\alpha_j+1$.  It
is apparent that the entire integer sets $\{1,2,...,2\alpha_i\}$
and $\{1,2,...,2\alpha_j\}$ cannot be used for unequal repunits.
The number of sequences satisfying the congruence relations for
each repunit must equal $\sum_{\sum t_k = 2\alpha_i+1}(2^{t_k}-1)$
and $\sum_{\sum t_l^\prime =2\alpha_j+1}(2^{t_l^\prime}-1)$, with
sequences of integers which sum to a non-zero value giving rise to
cancellation of complex numbers in the product being included.

Consider two products $\prod_{m=1}^{t_k}
(q_i-\omega_{2\alpha_i+1}^{s_m})$ and $\prod_{n=1}^{{t^\prime}_k}
(q_j-\omega_{2\alpha_j+1}^{s_n^\prime})$.  Upper and lower bounds
for these products are
$$\eqalign{(q_i-1)^{t_k}&< \prod_{m=1}^{t_k}
 (q_i-\omega_{2\alpha_i+1}^{s_m}) <
(q_i+1)^{t_k} \cr (q_j-1)^{t_l^\prime}& < \prod_{n=1}^{t_l^\prime}
(q_j-\omega_{2\alpha_j+1}^{s_n^\prime}) < (q_j+1)^{t_l^\prime} \cr}
\eqno(2.11)
$$
Since the minimum difference between primes is 2, one of the
inequalities $(q_i+1)^{t_k} \le (q_j-1)^{t_l^\prime}$,
$(q_j+1)^{t_l^\prime} \le (q_i-1)^{t^k}$ $(q_i+1)^{t_k} >
(q_j-1)^{t_l^\prime}$ or  $(q_j+1)^{t_l^\prime} > (q_i-1)^{t^k}$
will be satisfied when $t_l^\prime\ne t_k$.  The first two
inequalities imply that the factors of the repunits defined by the
products $\prod_{m=1}^{t_k} (q_i-\omega_{2\alpha_i+1}^{s_m})$ and
$\prod_{n=1}^{t_k^\prime} (q_j-\omega_{2\alpha_j+1}^{s_n^\prime})$
are distinct.

If either of the next two inequalities hold, the factors possibly
could be equal when $(q_i-1)^{t_k}=(q_j-1)^{t_i^\prime}$ or
$(q_i+1)^{t_k}=(q_j+1)^{t_l^\prime}$. In the first case,
$t_l^\prime=n_{ij}t_k$, $q_i-1=(q_j-1)^{n_{ij}},~n_{ij}>2$.  This
relation holds for all $t_k$ and $t_l^\prime$ such that $\sum_k
t_k=n_{ij} \sum_l t_l^\prime$.  However, $2\alpha_i+1$ and
$2\alpha_j+1$ are prime, so that this equation is not valid. When
$q_i+1=(q_j+1)^{n_{ij}},~n_{ij}\ge 2$, the relation $\sum_k t_k=
n_{ij} \sum_l t_l^\prime$ again leads to a contradiction.

The identification of the products in equation (2.11) for each
corresponding $k,l$ is necessary, since any additional product would
give rise either to a new prime factor after appropriate rescaling,
or a different power of the same number, which would would prevent a
cancellation of an extra prime, described later in the proof. The
inequalities $(q_j-1)^{t_l^\prime} < (q_i-1)^{t_k} <
(q_j+1)^{t_l^\prime}$ imply that
$$t_l^\prime~{{ln(q_j-1)}\over {ln(q_i-1)}}<t_k < t_l^\prime
{{ln(q_j+1)}\over {ln(q_i-1)}} \eqno(2.12)
$$
If these inequalities hold for two pairs of exponents
$(t_{k_1},t_{l_1}^\prime)$, $(t_{k_2},t_{l_2}^\prime)$, the interval
\hfil\break $\left[t_{l_2}^\prime {{ln(q_j-1)}\over {ln(q_i-1)}},
t_{l_2}^\prime {{ln(q_j+1)}\over {ln(q_i-1)}}\right]$ contains
${{t_{l_2}^\prime}\over {t_{l_1}^\prime}}t_{k_1}$.  The fractions
${{t_{k_m}}\over {t_{k_n}}}$ cannot be equal to
${{t_{l_m}^\prime}\over {t_{l_n}^\prime}}$ for all $m,n$ as this
would imply that ${{t_k}\over {t_l^\prime}}={{2\alpha_i+1}\over
{2\alpha_j+1}}$, which is not possible since ${{2\alpha_i+1}\over
{2\alpha_j+1}}$ is an irreducible prime fraction when $\alpha_i\ne
\alpha_j$.

Then
$$\left\vert {{t_{k_1}}\over {t_{l_1}^\prime}}-{{t_{k_2}}\over
 {t_{l_2}^\prime}}
\right\vert\ge {1\over {t_{l_1}^\prime t_{l_2}^\prime}} \eqno(2.13)
$$
Then $t_{k_2}\not\in \left[t_{l_2}^\prime {{ln(q_j-1)}\over
{ln(q_i-1)}}, t_{l_2}^\prime {{ln(q_j+1)}\over {ln(q_i-1)}}\right]$
if
$$ln~\left({{q_j+1}\over {q_j-1}}\right) < {{ln(q_i-1)}\over
 {t_{l_1}^\prime
t_{l_2}^\prime}} \eqno(2.14)
$$
As $t_l^\prime \le 2\alpha_j+1$, this inequality is valid when
$$q_j\cdot ln(q_i-1) > 3(2\alpha_j+1)^2
\eqno(2.15)
$$

Since the arithmetic primitive factor of ${{q^n-1}\over {q-1}},~n\ne
6$ is $\Phi_n(q)=\prod_{{1\le k\le n}\atop
{gcd(k,n)=1}}(q_i-e^{{2\pi i k}\over n})$ or ${{\Phi_n(q)}\over p}$,
where $p$ is the largest prime factor of both ${n\over {gcd(n,3)}}$
and $\Phi_n(q)$ [1][2][4][32][35], any prime, which is a primitive
divisor of ${{q_i^{2\alpha_i+1}-1}\over {q_i-1}}$, divides
$\Phi_{2\alpha_i+1}(q_i)$. It follows that these primes can be
obtained by appropriate multiplication of the products
$\prod_{m=1}^{t_k} (q_i-\omega_{2\alpha_i+1}^{s_m})$   and
$\prod_{n=1}^{t_k^\prime}(q_j-\omega_{2\alpha_j+1}^{s_n^\prime}) $.
Although these products may not be the primitive divisors, they are
real, and multiplication of $\prod_{m=1}^{t_1}
(q_i-\omega_{2\alpha_i+1}^{s_m})$ by $\kappa_1\in {\bf R}$ can be
compensated by $\prod_{m^\prime=1}^{t_2}
(q_i-\omega_{2\alpha_i+1}^{s_m^\prime})$  by $\kappa_1^{-1}$ to
obtain the integer factor.  Since products of two complex linear
factors $(q_i-e^{{2\pi i k_1}\over {2\alpha_i+1}})(q_i-e^{{-2\pi i
k_1}\over {2\alpha_i+1}})$ are real, let
$$\eqalign{x_1&=(q_i-e^{{2\pi ik_1}\over {2\alpha_i+1}})(q_i-e^{-{{2\pi
 i k_1}
\over {2\alpha_i+1}}}) \cr x_2&=(q_i-e^{{2\pi ik_2}\over
{2\alpha_i+1}})(q_i-e^{-{{2\pi i k_2} \over {2\alpha_i+1}}}) \cr
y_1&=(q_j-e^{{2\pi ik_3}\over {2\alpha_j+1}})(q_i-e^{-{{2\pi i k_3}
\over {2\alpha_j+1}}}) \cr y_2&=(q_j-e^{{2\pi ik_4}\over
{2\alpha_j+1}})(q_i-e^{-{{2\pi i k_4} \over {2\alpha_j+1}}}) \cr}
\eqno(2.16)
$$
The occurrence of the same prime divisors in the two repunits
${{q_i^{2\alpha_i+1}-1} \over {q_i-1}}$ and
${{q_j^{2\alpha_j+1}-1}\over {q_j-1}}$ yields relations of the
form
$$\eqalign{
\kappa_1x_1&= P_1^{\ell_1}=P_1^{h_1}\kappa_1^\prime y_1 \cr
\kappa_1^{-1} x_2 &=P_2^{\ell_2}= P_2^{h_2}\kappa_1^{\prime -1} y_2
\cr} \eqno(2.17)
$$
for two prime factors implies
$$P_1^{\ell_2 h_1-h_2 \ell_1}={{x_1^{\ell_2-h_2}
 x_2^{\ell_2-{{h_1\ell_2}\over
{\ell_1}}}}\over {y_1^{\ell_2}y_2^{\ell_2}}} \eqno(2.18)
$$
which cannot be satisfied by the prime $P_1$.  For three prime
factors,
$$\eqalign{
\kappa_1 x_1&=P_1^{\ell_1}= P_1^{h_1} \kappa_1^\prime y_1 \cr
\kappa_1^{-1}\kappa_2 x_2&=P_2^{\ell_2}= P_2^{h_2} \kappa_1^{\prime
-1} \kappa_2^\prime y_2 \cr \kappa_2^{-1}
x_3&=P_3^{\ell_3}=P_2^{h_3} \kappa_2^{\prime -1} y_3 \cr}
\eqno(2.19)
$$
yielding the equation
$$P_1^{\ell_2 h_1-h_2 \ell_1} P_3^{\ell_2 h_3-h_2 \ell_3}
={{(x_1x_2 x_3)^{\ell_2-h_2}}\over {y_1y_2y_3}} \eqno(2.20)
$$
However, ${{(x_1x_2)^{\ell_2-h_2}}\over {y_1y_2}}$ introduces a
minimum of two different primes, whereas ${{x_3^{\ell_2-h_2}}\over
{y_3}}$ contains at least one more prime divisor.  Alternatively,
this fraction equals
$${{(x_1x_2)^{{(\ell_2-h_2)}\over 2}}\over {(y_1y_2)^{1\over 2}}}
\cdot {{(x_2x_3)^{{(\ell_2-h_2)}\over 2}}\over {(y_2y_3)^{1\over
2}}} \cdot {{(x_3x_1)^{{(\ell_2-h_2)}\over 2}}\over
{(y_3y_1)^{1\over 2}}}\eqno(2.21)
$$
Each term in the product contains $n$ primes, $n\ge 2$, and
cancellation between the fractions leads to the deletion of a
maximimum of $n-1$ primes, since cancellation of all of the primes
would imply that the ratio of a power of one term and the other term
is a prime power, which is not possible based on the form of the
linear complex factors.  Consequently, the product (2.21) introduces
at least three distinct primes.  Continuing this factorization for
products of more than three prime powers, it follows that there is a
distinct prime divisor amongst the factors of $q_i^{2\alpha_i+1}-1$
and $q_j^{2\alpha_j+1}-1$, with $q_i,q_j, 2\alpha_i+1, 2\alpha_j+1$
being odd primes.

Since the prime divisors of the repunits can be obtained by
rescaling of products of the form (2.11), $t_{l_1}^\prime$ and
$t_{l_2}^\prime$ can be set equal, provided that the factor of a
power of a prime is included to match the intervals.  Then,
$P^h(q_j-1)^2 < (q_i-1)^{t_{k_1}} < P^h(q_j+1)^2$ and $t_{k_1}\in
\left[{{2ln(q_j-1)+h~ln P}\over {ln(q_i-1)}},
{{2ln(q_j+1)+h~lnP}\over {ln(q_i-1)}}\right]$.  Again, for some
$k_1$, $k_2$, ${{t_{k_1}}\over {t_{l_1}^\prime}}\ne {{t_{k_2}}\over
{t_{l_2}^\prime}}$ if $\alpha_i\ne \alpha_j$, and $\vert
t_{k_1}-t_{k_2}\vert \ge 1$.  The inequalities $P^h(q_j-1)^2 <
(q_i-1)^{t_{k_2}} < P^h(q_j+1)^2$ arise because each of the terms
$y_l=(q_j-e^{{2\pi i k_l}\over {2\alpha_j+1}}) (q_j-e^{{2\pi i
k_l}\over {2\alpha_j+1}})$, which is matched with a corresponding
product of linear factors
$\prod_{s_m=1}^{t_k}~(q_i-\omega_{2\alpha_i+1}^{s_m})$, has the same
upper and lower bound.  However, $t_{k_2}\not\in
\left[{{2ln(q_j-1)+h~ln P}\over {ln(q_i-1)}},{{2ln(q_j+1)+h~ln
P}\over {ln(q_i-1)}}\right]$ if ${{2ln\left({{q_j+1}\over
{q_j-1}}\right)}\over {ln(q_i-1)}}< 1$ which is implies $q_i> 5$.
The equality is also valid because $t_{k_2}={{2ln(q_j+1)+h~ln
P}\over {ln(q_i-1)}}$ only if $P^h={{(q_i-1)^{t_{k_2}}}\over
{(q_j+1)^2}}$ which implies that
$q_i-1=(q_j+1)^{h^\prime},~h^\prime\ge 1$. Then $t_l^\prime=h^\prime
t_k$, and if $h^\prime > 1$, summation over $k,l$ gives
$2\alpha_j+1=h^\prime(2\alpha_i+1)$ which is not possible as
$2\alpha_i+1$ and $2\alpha_j+1$ are prime. If $h^\prime=1$,
$t_k=t_l^\prime$ and $2\alpha_i+1$ must be set equal to
$2\alpha_j+1$.  Setting $q_j=3$ and $q_i=5$, it can be proven by
induction that one of the repunits ${{3^{2\alpha_i+1}-1}\over 2}$
and ${{5^{2\alpha_i+1}-1}\over 4}$ has a distinct prime divisor.
First, ${{3^3-1}\over 2}=13$, ${{3^5-1}\over 2}=11\cdot 11$,
${{5^3-1}\over {5-1}}=31$ and ${{5^5-1}\over 4}=11\cdot 71$.
Secondly, when $2\alpha_i+1$, there are only primitive divisors.

For the known solutions to the equation ${{x^m-1}\over
{x-1}}={{y^n-1}\over {y-1}}$, one of the prime bases is 2.  Since
$q_j-1=1$, the theorem is circumvented because the bounds (2.11) are
satisfied for a larger set of primes $q_i$ and exponents $t_k$.
Furthermore, the integers sets $\{1,2,...,2\alpha_i\}$ and
$\{1,2,...,2\alpha_j\}$ are used entirely in the products to give
the same integer, which must then be a prime. The existence of
solutions to the inequalities $1<(q_i-1)^{2\alpha_i} <
3^{2\alpha_i}$ implies that these bounds do not exclude the equality
of ${{2^{2\alpha_j+1}-1}\over {q_j-1}}$ and
${{q_i^{2\alpha_i+1}-1}\over {q_i-1}}$ for some
$2\alpha_j+1,q_i,2\alpha_i+1$.

If the primes which divide only $q_i^{2\alpha_i+1}-1$, are factors
of $q_i-1$ or $q_j-1$, they would be partially cancelled in a
comparison of ${{q_i^{2\alpha_i+1}-1}\over {q_i-1}}$ and
${{q_j^{2\alpha_j+1}-1}\over {q_j-1}}$.  Any divisor of $q_i-1$
which is a factor of ${{q_i^{2\alpha_i+1}-1}\over {q_i-1}}$ also
must divide $2\alpha_i+1$.  When $2\alpha_i+1$ and $2\alpha_j+1$
are prime, this divisor would have to be $2\alpha_i+1$ or
$2\alpha_j+1$.

An exception to the result concerning the occurrence of a distinct
prime factor in one of the repunits could occur if equations of
the form
$$\eqalign{{{q_i^{2\alpha_i+1}-1}\over {q_i-1}}~&=~(2\alpha_i+1)^\nu
                                         {{q_j^{2\alpha_j+1}-1}\over
 {q_j-1}}
\cr (2\alpha_j+1)^\nu {{q_i^{2\alpha_i+1}-1}\over {q_i-1}}~&=~
                             {{q_j^{2\alpha_j+1}-1}\over {q_j-1}}
\cr (2\alpha_j+1)^{\nu_1}{{q_i^{2\alpha_i+1}-1}\over {q_i-1}}~&=~
   (2\alpha_i+1)^{\nu_2}{{q_i^{2\alpha_i+1}-1}\over {q_i-1}}
\cr} \eqno(2.22)
$$
are satisfied, as the set of prime divisors of the repunits would be
identical if $2\alpha_i+1\bigg\vert {{q_j^{2\alpha_j+1}-1}\over
{q_j-1}}$ in the first relation, $2\alpha_j+1\bigg\vert
{{q_i^{2\alpha_i+1}-1}\over {q_i-1}}$ in the second condition and
$2\alpha_i+1\bigg\vert {{q_j^{2\alpha_j+1}-1}\over {q_j-1}},~
2\alpha_j+1\bigg\vert {{q_i^{2\alpha_i+1}-1}\over {q_i-1}}$ in the
third relation. The conditions with $\nu=\nu_1=\nu_2=1$ follow if
different powers of the other primes do not arise.  However,
$q_i^p-1\equiv 0~(mod~p^\nu)$ only if $q_i^p-1\equiv 0~(mod~p)$, and
since $q_i^p-q\equiv 0~(mod~p)$,  this is possible when $p\vert
q_i-1$. If $p\vert q_i-1$, then ${{q_i^p-1}\over {q_i-1}}\equiv p$
and ${{q_i^p-1}\over {q_i-1}}\not\equiv 0~(mod~p^\nu),~\nu\ge 2$.
Setting $\nu=\nu_1=\nu_2$ gives
$$\eqalign{{{q_i^{2\alpha_i+1}-1}\over {q_i-1}}~&=~(2\alpha_i+1)
                                         {{q_j^{2\alpha_j+1}-1}\over
 {q_j-1}}
\cr (2\alpha_j+1){{q_i^{2\alpha_i+1}-1}\over {q_i-1}}~&=~
{{q_j^{2\alpha_j+1}-1}\over {q_j-1}}
\cr (2\alpha_j+1){{q_i^{2\alpha_i+1}-1}\over {q_i-1}}~&=~
   (2\alpha_i+1){{q_i^{2\alpha_i+1}-1}\over {q_i-1}}
\cr} \eqno(2.23)
$$
with $2\alpha_i+1\not\vert {{q_j^{2\alpha_j+1}-1}\over {q_j-1}}$
and $2\alpha_j+1\not\vert {{q_i^{2\alpha_i+1}-1}\over {q_i-1}}$.
It follows that either $2\alpha_i+1$ or $2\alpha_j+1$ is a prime
which divides only one of the repunits.

Furthermore, $q_i-1$ and $q_j-1$ are integers which are not
rescaled, such that distinct prime divisors arise in the products of
the remaining linear factors. If the exponents $2\alpha_i+1$ and
$2\alpha_j+1$ are unequal, they represent factors which do not
divide both repunits.  When the exponents $2\alpha_i+1$ and
$2\alpha_j+1$ are equal, it would be necessary for $n_{ij}$ to be
equal to one, which is not feasible since $q_i\ne q_j$. It follows
also that ${{q_i^{2\alpha_i+1}-1}\over {q_i-1}}$ and
${{q_j^{2\alpha_j+1}-1}\over {q_j-1}}$ cannot be equal, and the
original assumption of their inequality is valid.  However, by the
first two relations in Eq.(2.23), there exists a prime which
is not a divisor of both repunits.

 \line{\hfil{\qed}}

\hfil\break\hfil\break
\noindent
{\bf Example 2.}
\vskip 5pt
\noindent
An example of congruence with more than one solution less than $P$
is
$$\eqalign{
(2\alpha_i+1)&+\alpha_i(2\alpha_i+1)x+{{(2\alpha_i+1)2\alpha_i(2\alpha_i-1)}
\over {3!}}x^2+{{(2\alpha_i+1)2\alpha_i(2\alpha_i-1)(2\alpha_i-2)}
\over {4!}}x^3 \cr
&+{{(2\alpha_i+1)2\alpha_i(2\alpha_i-1)(2\alpha_i-2)(2\alpha_i-3)}\over
{5!}}x^4 =0~(mod~11) \cr} \eqno(2.24)
$$
which is solved by $x=2,~4$ when $2\alpha_i+1=5$.   This is
consistent with the inequality (2.6), since $\epsilon=0.074897796$
when $x_0=2$. Consequently, ${{3^5-1}\over 2}=11\cdot 11$ and
${{5^5-1}\over 4}=11\cdot 71$ both have the divisor 11 and the
distinct prime divisor arises in the larger repunit. Indeed,
${{3^5-1}\over 2}$ would not have a different prime factor from a
repunit ${{q_k^{2\alpha_k+1}-1}\over {q_k-1}}$ that has $11$ as a
divisor. A set of primes $\{3,q_k,...\}$ and exponents
$\{5,2\alpha_k,...\}$ does not represent an exception to the Theorem
1 if ${{3^5-1}\over 2}$ is chosen to be the first repunit in the
product, introducing the prime divisor 11.  Each subsequent repunit
then would have a distinct prime factor from the previous repunit in
the sequence.

\line{\hfil{$\square$}}

\vskip 10pt\noindent{\bf 3. Formulation of the
Condition for Perfect Numbers in terms of the} \hfil\break
\phantom{.....}{\bf Coefficients of Repunits} \vskip 10pt \noindent
Let $N=(4k+1)^{4m+1}\prod_{i=1}^\ell~q_i^{2\alpha_i}$ [13] and the
coefficients $\{a_i\}$ and $\{b_i\}$ be defined by
$$a_i{{q_i^{2\alpha_i+1}-1}\over {q_i-1}}=
b_i{{(4k+1)^{4m+2}-1}\over {4k}}~~~~~~~~~ gcd(a_i,b_i)=1
\eqno(3.1)
$$
If ${{\sigma(N)}\over N}\ne 2$,
$$\eqalign{
{\sqrt{2(4k+1)}}&\left[{{q_1^{2\alpha_1+1}-1}\over {q_1-1}}
{{q_2^{2\alpha_2+1}-1} \over
{q_2-1}}...{{q_\ell^{2\alpha_\ell+1}-1}\over {q_\ell-1}}
{{(4k+1)^{4m+2}-1}\over {4k}}\right]^{1\over 2} \cr
&~~~~~~~~~~~={\sqrt{2(4k+1)}}{{(b_1...b_\ell)^{1\over 2}}\over
{(a_1...a_\ell)^{1\over 2}}} \cdot \left({{(4k+1)^{4m+2}-1}\over
{4k}}\right)^{{(\ell+1)}\over 2} \cr &~~~~~~~~~~~\ne
2(4k+1)^{2m+1}\prod_{i=1}^\ell~q_i^{\alpha_i} \cr} \eqno(3.2)
$$
or
$$\eqalign{
{{b_1...b_\ell}\over {a_1...a_\ell}}&=\prod_{i=1}^\ell~
{{q_i^{2\alpha_i+1}-1} \over {q_i-1}}\cdot \left({{4k}\over
{(4k+1)^{4m+2}-1}}\right)^\ell \cr &\ne 2(4k+1)^{4m+1}
\left[{{4k}\over {(4k+1)^{4m+2}-1}}\right]^{\ell+1}
\prod_{i=1}^{\ell-1}~q_i^{2\alpha_i}\cdot
{{q_\ell^{2\alpha_\ell+1}-1}\over {q_\ell-1}} \cr} \eqno(3.3)
$$
When $\ell > 5$ is odd, there exists an odd integer $\ell_o$ and
an even integer $\ell_e$ such that $\ell=3\ell_0+2\ell_e$, so that
$$\eqalign{
{{b_1...b_\ell}\over {a_1...a_\ell}}&=\left({{b_{13}}\over {a_{13}}}
{{a_2}\over {b_2}}\right) \left({{b_{46}}\over {a_{46}}}{{a_5}\over
{b_5}}\right)... \left({{b_{3\ell_o-2, 3\ell_o}}\over {a_{3\ell_o-2,
3\ell_o}}} {{a_{3\ell_o-1}}\over
{b_{3\ell_o-1}}}\right)\left({{b_{3\ell_o+1}b_{3\ell_o+2}} \over
{a_{3\ell_o+1}a_{3\ell_o+2}}}\right) \cr &~~~~~~~~~~~~~~~~~~~~~~~
... \left({{b_{\ell-1}b_\ell}\over {a_{\ell-1}a_\ell}}\right)\cdot
{{s^2}\over {t^2}} \cr} \eqno(3.4)
$$
with $s,~t\in {\Bbb Z}$. It has been proven that ${{b_{3{\bar i}-2,
3{\bar i}}}\over {a_{3{\bar i}-2, 3{\bar i}}}}{{a_{3{\bar
i}-1}}\over {b_{3{\bar i}-1}}}\ne 2(4k+1)\cdot {{s^2}\over {t^2}}$
for any choice of $a_{3{\bar i}-2}, a_{3{\bar i}-1}, a_{3{\bar i}}$ and
$b_{3{\bar i}-2}, b_{3{\bar i}-1}, b_{3{\bar i}}$ consistent with
Eq.(3.1) [11]. Similarly, ${{b_\ell}\over {a_\ell}}\ne
2(4k+1)\cdot {{s^2}\over {t^2}}$ so that
$$\eqalign{{{b_{3{\bar i}-2, 3{\bar i}}}\over {a_{3{\bar i}-2, 3{\bar
 i}}}}
{{a_{3{\bar i}-1}}\over {b_{3{\bar i}-1}}}&\equiv 2(4k+1){{{\bar
\rho}_{3{\bar i}-2}} \over {{\bar \chi}_{3{\bar i}-2}}}\cdot
{{s^2}\over {t^2}} \cr {{b_{3{\bar j}+1}b_{3{\bar j}+2}}\over
{a_{3{\bar j}+1}a_{3{\bar j}+2}}} &=2(4k+1){{{\hat \rho}_{3{\bar
j}+2}}\over {{\hat \chi}_{3{\bar j}+2}}}\cdot {{s^2}\over {t^2}}
\cr} \eqno(3.5)
$$
where the fractions are square-free and $gcd({\bar \rho}_{3{\bar
i}-2}, {\bar \chi}_{3{\bar i}-2})=1$, $gcd({\hat \rho}_{3{\bar
j}+2}, {\hat \chi}_{3{\bar j}+2})=1$. \footnote{*}{The notation
has been changed from that of reference [11] with a different
choice of \phantom{.....}index for ${\bar \rho}$, ${\bar \chi}$.}
Then,
$${{b_1...b_\ell}\over {a_1...a_\ell}}=2(4k+1)\cdot
{{{\bar \rho}_1}\over {{\bar \chi}_1}}...{{{\bar
\rho}_{3\ell_o-2}}\over {{\bar \chi}_{3\ell_o-2}}}  {{{\hat
\rho}_{\ell-2\ell_e+2,2}}\over {{\hat
\chi}_{\ell-2\ell_e+2,2}}}...{{{\hat \rho}_{\ell 2}}\over {{\hat
\chi}_{\ell 2}}}\cdot {{s^2}\over {t^2}} \eqno(3.6)
$$
When $\ell=2\ell_o+3\ell_e > 4$ for odd $\ell_0$ and even
$\ell_e$,
$${{b_1...b_\ell}\over
 {a_1...a_\ell}}=2(4k+1)\left({{4k+1)^{4m+2}-1}\over {4k}}
\right)\cdot {{{\hat \rho}_{22}}\over {{\hat \chi}_{22}}}... {{{\hat
\rho}_{2\ell_o,2}}\over {{\hat \chi}_{2\ell_o,2}}} {{{\bar
\rho}_{\ell-3\ell_e+1}}\over {{\bar \chi}_{\ell-3\ell_e+1}}}
...{{{\bar \rho}_{\ell-2}}\over {{\bar \chi}_{\ell-2}}}\cdot
{{s^2}\over {t^2}} \eqno(3.7)
$$
If the products
$$\prod_{{\bar i}=1}^{\ell_o} \left({{{\bar \rho}_{3{\bar i}-2}}\over
{{\bar \chi}_{3{\bar i}-2}}}\right) \prod_{{\bar j}=1}^{\ell_e}
\left({{{\hat \rho}_{3\ell_o+2{\bar j},2}}\over {{\hat
\chi}_{3\ell_o+2{\bar j},2}}}\right) \eqno(3.8)
$$
for odd $\ell$ and
$$\prod_{{\bar i}=1}^{\ell_o}\left(
{{{\hat \rho}_{2{\bar i},2}}\over {{\hat \chi}_{2{\bar
i},2}}}\right) \prod_{j=1}^{\ell_e-1}\left({{{\bar
\rho}_{2\ell_o+3{\bar j}+1}}\over {{\bar \chi}_{2\ell_o+3{\bar
j}+1}}}\right) \eqno(3.9)
$$
for even $\ell$ are not the squares of rational numbers,
$$\eqalign{{{b_1...b_\ell}\over {a_1...a_\ell}}&\ne 2(4k+1)\cdot
 {{s^2}\over {t^2}}~~~~~~~~~~~
~~~~~~~\ell~is~odd \cr {{b_1...b_\ell}\over {a_1...a_\ell}}&\ne
2(4k+1) \left({{4k+1)^{4m+2}-1}\over {4k}} \right)\cdot {{s^2}\over
{t^2}}~~~~~~~~~\ell~is~even \cr} \eqno(3.10)
$$
which would imply the inequality (3.3) and the non-existence of an
odd perfect number. \vskip 10pt \noindent {\bf 4. Examples of
Integers satisfying the Rationality Condition} \vskip 10pt
\noindent A set of integers
$N=(4k+1)^{4m+1}\prod_{i=1}^\ell~q_i^{2\alpha_i}$ which satisfy
the condition of rationality of
$$\left[2(4k+1){{q_1^{2\alpha_1+1}-1}\over {q_1-1}}...
{{q_\ell^{2\alpha_\ell+1}-1}\over {q_\ell-1}} {{(4k+1)^{4m+2}-1}
\over {4k}}\right]^{1\over 2} \eqno(4.1)
$$
is given in the following list:
$$\eqalign{& 37^2 3^2 5^2 29^2 79^2 83^2 137^2 283^2 313^2
\cr & 37^5 3^2 29^2 67^2 79^2 83^2 137^2 283^2
\cr & 37^5 3^2 7^2 11^2 29^2 79^2 83^2 137^2 191^2 283^2
\cr & 37^5 3^2 5^2 11^2 13^2 29^2 47^2 79^2 313^2
\cr & 37^5 11^2 29^2 79^2 211^2 313^2
\cr & 37^5 13^2 29^2 47^2 79^2 83^2 137^2 211^2 283^2 313^2
\cr & 37^5 5^2 7^2 11^2 13^2 29^2 47^2 79^2 83^2 137^2 191^2 211^2
 283^2
\cr & 37^5 3^2 11^2 13^2 29^2 47^2 67^2 79^2
\cr & 37^5 3^2 7^2 13^2 29^2 47^2 79^2 191^2 313^2
\cr & 37^5 7^2 11^2 13^2 29^2 47^2 67^2 79^2 83^2 137^2 191^2 211^2
 283^2 313^2
\cr & 37^5 5^2 11^2 29^2 67^2 79^2 211^2
\cr & 37^5 5^2 7^2 13^2 29^2 47^2 79^2 191^2 211^2 313^2 \cr}
 \eqno(4.2)
$$
None of these integers satisfy the condition ${{\sigma(N)}\over
N}=2$.  For example, the sum of divisors of the first integer is
$$\sigma(37^5 3^2 5^2 29^2 79^2 83^2 137^2 283^2 313^2)=2\cdot 37\cdot
 3^4\cdot 7^4
\cdot 13^2 \cdot 19^2\cdot 31^2\cdot 43^2 \cdot 67^2 \cdot 73^2
\cdot 181^2 \cdot 367^2 \eqno(4.3)
$$
such that new prime factors $7,13, 19, 31, 43, 67, 73, 181, 367$ are
introduced.  The inclusion of these prime factors in the integer
$N$ leads to yet additional primes, and then the lack of closure
of the set of prime factors renders it impossible for them to be
paired to give even powers in $\sigma(N)$ with the exception of
$2(4k+1)$.

It may be noted that the decompositions of repunits with prime
bases of comparable magnitude and exponent 6 include factors that
are too large and cannot be matched easily with factors of other
repunits. It also can be established that repunits with prime
bases and other exponents do not have square-free factors that can
be easily matched to provide a closed sequence of such pairings.
This can be ascertained from the partial list
$$\eqalign{U_5(6,5)&={{5^5-1}\over 4}=781
\cr U_6(6,5)&=3906=2\cdot 7 \cdot 31\cdot 3^2 \cr U_5(8,7)&=2801
\cr U_5(12,11)&=16105 \cr U_5(14,13)&=30941 \cr U_5(18,17)&=88741
\cr U_7(4,3)&=1093 \cr U_7(6,5)&=19531 \cr U_9(4,3)&=9841 \cr
U_3(32,31)&=993=3\cdot 331 \cr} \eqno(4.4)
$$
The entire set of integers satisfying the rationality condition
is therefore restricted to a ring of prime bases and powers
and exponents $4m+1=5$, $2\alpha_i=1$ with a given set of
prime factors occurring in the sum-of-divisors function.

The first prime of the form $4k+1$ which arises as a coefficient
$D$ in the equality
$${{q^3-1}\over {q-1}}=Dy^2~~~~~~~~~~~~~~q~prime
\eqno(4.5)
$$
is 3541.  This prime is too large to be the basis for the factor
${{(4k+1)^{4m+2}-1}\over {4k}}$, since it would give rise to many
other unmatched prime factors in the product of repunits and the
rationality condition would not be satisfied.  Amongst the
coefficients $D$ which are composite, the least integer with a
prime divisor of the form $4k+1$ is 183, obtained when $q=13$.
However, the repunit with base 61 still gives rise to factors
which cannot be matched since
$${{61^6-1}\over {60}}=858672906=2\cdot 3\cdot 7 \cdot 13 \cdot 31
 \cdot 97 \cdot 523
\eqno(4.6)
$$
Therefore, the rationality condition provides confirmation of
the nonexistence of odd perfect numbers, which, however, can
be proven with certainty only through the methods described in this
work.

\vskip 10pt \noindent {\bf 5. On the Non-Existence of Coefficients
of Repunits satisfying the} \hfil\break \noindent
\phantom{.....}{\bf Perfect Number Condition} \vskip 10pt
\noindent

Based on Theorem 1, it is proven that additional prime divisors
arise in the sum-of-divisors function and a relation
equivalent to the perfect number condition cannot be
satisfied by the primes $4k+1$ and $q_i,~i=1,...,\ell$.
\hfil\break\hfil\break
\noindent
{\bf Theorem 2.} There does not exist any set of odd primes
$\{4k+1;q_1,...,q_\ell\}$ such that the coefficients $\{a_i\}$ and
$\{b_i\}$ satisfy the equation
$${{b_1...b_\ell}\over {a_1...a_\ell}}= 2(4k+1)
                    \left[{{4k}\over {(4k+1)^{4m+2}-1}}\right]^{\ell+1}
                               \prod_{i=1}^\ell q_i^{2\alpha_i}.
$$
\hfil\break \hfil\break {\bf Proof.} It has been observed that
${{b_1}\over {a_1}}\ne 2(4k+1)\cdot {{s^2}\over {t^2}}$ by the
non-existence of multiply perfect numbers with less than four prime
factors [8][9],
$${{b_1b_2}\over {a_1a_2}}\ne 2(4k+1)\left[{{4k}\over
 {(4k+1)^{4m+2}-1}}\right]^3
  q_1^{2\alpha_1}q_2^{2\alpha_2}
\eqno(5.1)
$$
by the non-existence of perfect numbers with three prime divisors
[10], and proven that ${{b_1b_2b_3}\over {a_1a_2a_3}}\ne
2(4k+1)\cdot {{s^2}\over {t^2}}$ so that the inequality is valid for
$\ell=1,2,3$.

Suppose that there are no odd integers $N$ of the form
$(4k+1)^{4m+1}\prod_{i=1}^{\ell-1} q_i^{2\alpha_i}$ with
${{\sigma(N)}\over N}=2$ so that
$${{b_1...b_{\ell-1}}\over {a_1...a_{\ell-1}}}\ne
2(4k+1)\left[{{4k}\over {(4k+1)^{4m+2}-1}}\right]^\ell
\prod_{i=1}^{\ell-1} q_i^{2\alpha_i} \eqno(5.2)
$$
If there exists a perfect number with prime factors
$\{4k+1,q_1,...,q_\ell\}$, then ${{b_1...b_\ell}\over
{a_1...a_\ell}}$ must have
$1+\ell+\tau\left({{(4k+1)^{4m+2}-1}\over
{4k}}\right)-\tau(U_{4m+2}(4k+2,4k+1), \prod_{i=1}q_i)$ distinct
prime factors, where $U_{4m+2}(4k+2,4k+1)$ is the Lucas number
${{(4k+1)^{4m+2}-1}\over {4k}}$ and
$\tau(U_{4m+2}(4k+2,4k+1),\prod_{i=1}q_i)$ denotes the number of
common divisors of the two integers. However, equality $\sigma(N)$
and 2N also implies that $\prod_{i=1}^\ell
{{q_i^{2\alpha_i+1}-1}\over {q_i-1}}$ has at least
$\ell+2-\tau\left({{(4k+1)^{4m+2}-1} \over {4k}}\right)$ different
prime divisors and a maximum of $\ell+1$ prime factors. If there
is no cancellation between ${{4k}\over {(4k+1)^{4m+2}-1}}$ and
$\prod_{i=1}^\ell~q_i^{2\alpha_i}$, multiplication of
$\prod_{i=1}^\ell U_{2\alpha_i+1}(q_i+1,q_i)=\prod_{i=1}^\ell
{{q_i^{2\alpha_i+1}-1}\over {q_i-1}}$  by
$U_{4m+2}(4k+2,4k+1)=\left({{4k}\over
{(4k+1)^{4m+2}-1}}\right)^\ell$ introduces $\tau\left({{4k}\over
{(4k+1)^{4m+2}-1}}\right)$ new prime divisors and
${{b_1...b_\ell}\over {a_1...a_\ell}}$ would have $\ell+2$
distinct prime factors. However, if $gcd(U_{4m+2}(4k+2,4k+1),
U_{2\alpha_i+1}(q_i+1,q_i))=1$ for all $i$, the repunit
$U_{4m+2}(4k+2,4k+1)$ must introduce additional prime divisors. It
follows that
$${{(4k+1)^{4m+2}-1}\over {8k}}=\prod_{{\bar i}\in I}~q_i^{2\alpha_i}
\eqno(5.3)
$$
must be valid for some integer set $I$.  There are no integer
solutions to
$${{x^n-1}\over {x-1}}=2y^2~~~~~~x\equiv 1(mod~4),~n\equiv 2(mod~4)
\eqno(5.4)
$$
and there is no odd integer satisfying these conditions. A variant
of this proof has been obtained by demonstrating the irrationality
of $\left[\prod_{i=1}^\ell~{{q_i^{2\alpha_i+1}-1}\over
{q_i-1}}\cdot \left({{8k(4k+1)}\over
{(4k+1)^{4m+2}-1}}\right)\right]^{1\over 2}$ when
$gcd(U_{2\alpha+1}(q_i+1,q_i), U_{4m+2}(4k+2,4k+1))=1$ [8].

When the number of prime factors of
$\prod_{i=1}^\ell~{{q_i^{2\alpha_i+1}-1}\over {q_i-1}}$ is less
than $\ell+1$, there must be at least two divisors of
$U_{4m+2}(4k+2,4k+1)$ which do not arise in the decomposition of
this product. While one of the factors is $2$, $4k+1$ is not a
divisor, implying that any other divisor must be $q_{\bar j}$ for
some ${\bar j}$. It has been assumed that there are no prime sets
$\{4k+1;q_1,...,q_{\ell-1}\}$ satisfying the perfect number
condition.    Either ${{b_1...b_{\ell-1}}\over
{a_1...a_{\ell-1}}}$ does not have $\ell+1$ factors, or if it does
have $\ell+1$ factors, then
$\prod_{i=1}^{\ell-1}~{{q_i^{2\alpha_i+1}-1}\over {q_i-1}}$ has
$\ell-1$ factors but
$$\prod_{i=1}^{\ell-1} {{q_i^{2\alpha_i+1}-1}\over {q_i-1}}\ne
 (4k+1)^{4m+1} \prod_{i\ne {\bar i},{\bar j}}
q_i^{2\alpha_i-t_i}
\eqno(5.5)
$$
for some prime $q_{\bar i}$, where $q_i^{t_i}\parallel
U_{4m+2}(4k+2,4k+1)$. Multiplication by
$U_{2\alpha_\ell+1}(q_\ell+1,q_\ell)$ must contain the factor
$q_{\bar i}^{2\alpha_{\bar i}}$, because ${{(4k+1)^{4m+2}-1}\over
{4k}}$ only would introduce the two primes $2,~q_{\bar j}$ and not
$q_{\bar i}$, since
$\prod_{i=1}^{\ell-1}U_{2\alpha_i+1}(q_i+1,q_i)$ contains $\ell-1$
primes, including $4k+1$ and excluding $q_{\bar i}$. Interchanging
the roles of the primes in the set $\{q_i, i=1,...,\ell\}$, it
follows that the prime equations
$$\eqalign{{{q_i^{2\alpha_i+1}-1}\over
 {q_i-1}}&=(4k+1)^{h_i}q_{j_i}^{2\alpha_{j_i}}
\cr {{q_\ell^{2\alpha_\ell+1}-1}\over {q_\ell-1}}&=(4k+1)^{h_\ell}
                                          q_{\bar i}^{2\alpha_{\bar i}}
\cr {{(4k+1)^{4m+2}-1}\over {4k}}&=2q_{\bar j}^{2\alpha_{\bar j}}
\cr} \eqno(5.6)
$$
must hold, where $h_\ell\ne 0$. Since the second equation has no
integer solution, $k\ge 1$ [11], it follows that
${{b_1...b_{\ell-1}}\over {a_1...a_{\ell-1}}}$ must not have
$\ell+1$ prime factors.

There cannot be less than $\ell+1$ different factors of
${{b_1...b_{\ell-1}}\over {a_1...a_{\ell-1}}}$, as each new
repunit ${{q_i^{2\alpha_i+1}-1}\over {q_i-1}}$ introduces at least
one distinct prime divisor by Theorem 1 and the factorization of
${{(4k+1)^{4m+2}-1}\over {4k}}$ contains at least two new primes.
Consequently, this would imply ${{b_1...b_{\ell-1}}\over
{a_1...a_{\ell-1}}}$ has at least $\ell+2$ prime factors and
${{b_1...b_\ell}\over {a_1...a_\ell}}$ has a minimum of $\ell+3$
prime divisors, which is larger than the number necessary for
equality between $U_{4m+2}(4k+2,4k+1)\prod_{i=1}^\ell
U_{2\alpha_i+1}(q_i+1,q_i)$ and $2(4k+1)\prod_{i=1}^\ell
q_i^{2\alpha_i}$.

If the maximum number of prime factors, $\ell+1$, is attained for
$\prod_{i=1}^\ell U_{2\alpha_i+1}(q_i+1,q_i)$, then the only
additional prime divisor arising from multiplication with
$U_{4m+2}(4k+2,4k+1)$ is 2.  However, since both 2 and $2k+1$
divide the repunit with base $4k+1$, this property does not hold
unless the prime factors of $2k+1$ and ${1\over
{2k+1}}\left({{(4k+1)^{4m+2}-1}\over {4k}}\right)$ can be included
in the set $\{q_i,i=1...,\ell\}$.  There are then a total of
$\ell+2$ prime factors in ${{b_1...b_\ell}\over {a_1...a_\ell}}$
only if $\left[{{8k}\over {(4k+1)^{4m+2}-1}}\right]^{\ell+1}$ can
be absorbed into $\prod_{i=1}^\ell q_i^{2\alpha_i}$.

The repunit then can be expressed as
$${{(4k+1)^{4m+2}-1}\over {4k}}=2 \prod_{j\in \{K\}} q_j^{t_j}
\eqno(5.7)
$$
where $\{K\}\subseteq \{1,...,q_\ell\}$. By the non-existence of
odd perfect numbers with prime factors $4k+1,q_i,i=1...,\ell-1$,
either
$U_{4m+2}(4k+2,4k+1)\prod_{i=1}^{\ell-1}U_{2\alpha_i+1}(q_i+1,q_i)$
has less than $\ell+1$ factors or
$$\prod_{i=1}^{\ell-1}{{q_i^{2\alpha_i+1}-1}\over {q_i-1}}\ne
 (4k+1)^{4m+1}
\prod_{i\in {\overline {\{K\}-\{q_{\bar i}\}}} }~q_i^{2\alpha_i}
 \prod_{\{K\}-\{q_{\bar i}\}} q_j^{2\alpha_j-t_j}
\eqno(5.8)
$$
for some ${\bar i}\in \{1,...,\ell\}$. If the product of repunits
for the prime basis $\{4k+1,q_1,...,q_{\ell-1}\}$ has less than
$\ell+1$ factors, ${{q_\ell^{2\alpha_\ell+1}-1}\over {q_\ell-1}}$
introduces at least two new prime factors. Even if one of these
divisors is $4k+1$, the other factor must be $q_{\bar i}$ for some
${\bar i}\ne \ell$. Interchange of the primes $q_i,i=1,...,\ell$
again yields the relations in equation (5.6).

If the product (5.8) includes $\ell$ primes, and not $q_{\bar i}$,
then $q_{\bar i}^{2\alpha_{\bar i}}$ must be a factor of
${{q_\ell^{2\alpha_\ell+1} -1}\over {q_\ell-1}}$.  Interchanging
the primes, it follows that the product (5.8) includes
$\prod_{i\ne {\bar i}}q_i^{2\alpha_i}$.

The prime divisors in $U_{2\alpha_\ell+1}(q_\ell+1,q_\ell)$ either
can be labelled $q_{j_\ell}$ for some $j_\ell\in \{1,...,\ell-1\}$
or equals $4k+1$.  while $4k+1$ does not divide
$U_{4m+2}(4k+2,4k+1)$, it can occur in the other repunits
${{q_i^{2\alpha_i+1}-1}\over {q_i-1}}$ for $i=1,...,\ell-1$.  The
prime $q_{j_\ell}=q_{\bar i}$ also may not be a divisor of
${{(4k+1)^{4m+2}-1}\over {4k}}$.

One choice for $q_\ell$ is a prime divisor of $2k+1$.  If $2k+1$ is
prime, the factors of ${{(2k+1)^{2\alpha_\ell+1}-1}\over {2k}}$ and
${{(4k+1)^{4m+2}-1}\over {4k}}$ can be compared.  Since the latter
quotient is equal to ${{(4k+1)^{2m+1}-1}\over {4k}}\cdot
[(4k+1)^{2m+1}+1]$, consider setting $m$ equal to $\alpha_\ell$.  By
Theorem 1, primitive divisor $P$ divides both repunits if $4k=nP+2k$
or $2k=nP$ implying $n=2$, $P=k$.  However, if $k\bigg\vert
{{(2k+1)^{2\alpha_\ell+1}-1}\over {2k}}$,
$1+(2k+1)+...+(2k+1)^{2\alpha_\ell}\equiv 2\alpha_\ell+1\equiv
0~(mod~k)$ and $k=2\alpha_\ell+1$ for prime exponents
$2\alpha_\ell+1$.  Suppose ${{(2k+1)^k-1}\over {2k}}=k\cdot
\prod_i~P_i^{m_i}$ and ${{(4k+1)^k-1}\over {4k}}=k\cdot
\prod_j~{\tilde P}_j^{n_j}$.  If a primitive divisor $rk+1$, $r\in
{\bf Z}$, divides both repunits, $(2k+1)^k-1=(rk+1)(x-1), ~x\in {\bf
Z}$, $(4k+1)^k-1=(rk+1)(x^\prime-1),~x^\prime\in {\bf Z}$, so that
$${{(2k+1)^k-1}\over {x-1}}={{(4k+1)^k-1}\over {x^\prime-1}}
\eqno(5.9)
$$
This relation would imply
$$x^\prime (2k+1)^k-(2k+1)^k-x^\prime= x(4k+1)^k-(4k+1)^k-x
\eqno(5.10)
$$
and therefore $x-1=b(4k+1)^k$, $x^\prime-1=a(4k+1)^k$.  However,
the equation is then
$$a(4k+1)^k(2k+1)^k-a(4k+1)^k~=~b(4k+1)^k(2k+1)^k-b(2k+1)^k
\eqno(5.11)
$$
which cannot be satisfied by any integers $a,b$.  Not all
primitive divisors of the two repunits are identical.  For the
exception, $k=1$, $2\alpha_\ell+1=5$, $4m+2=10n$,
${{(2k+1)^{2\alpha_\ell+1}-1}\over {2k}}=11^2$,
${{(4k+1)^{4m+2}-1}\over {4k}}$ has other prime factors in
addition to 11.

If $m\ne \alpha_\ell$, it can be demonstrated that there is a
primitive divisor of ${{(2k+1)^{2\alpha_\ell+1}-1}\over {2k}}$ which
is not a factor of ${{(4k+1)^{4m+2}-1}\over {4k}}$ by using the
comparison of the linear factors in the decomposition of each
numerator in Theorem 1, since $4k\ne (2k)^n,~n\ge 2$ and $4k+2\ne
(2k+2)^n,~n\ge 2$ for any $k>1$.

In general, either there exists a prime factor of
${{q_\ell^{2\alpha_\ell+1}-1}\over {q_\ell-1}}$ which does not
divide \hfil\break ${{(4k+1)^{4m+2}-1}\over {4k}}$ or there are more
than $\ell+2$ primes in the decomposition of the product of
repunits. The existence of a distinct prime divisor requires a
separate demonstration for composite exponents. Let $p_1(4m+2)$,
$p_2(4m+2)$ be two prime divisors of $4m+2$.  Then
${{q_\ell^{2\alpha_\ell+1}-1}\over {q_\ell-1}}$ has a prime factor
different from the divisors of ${{(4k+1)^{p_1(4m+2)}-1}\over {4k}}$
and ${{(4k+1)^{p_2(4m+2)}-1}\over {4k}}$ by Theorem 1. While the
union of the sets of prime divisors of the two repunits is contained
in the factorization of ${{(4k+1)^{p_1(4m+2)p_2(4m+2)}-1}\over
{4k}}$, the repunit with the composite exponent will have a
primitive divisor, which does not belong to the union of the two
sets and equals $a_1p_1p_2+1$. If this prime divides
${{q_\ell^{2\alpha_\ell+1}-1}\over {q_\ell-1}}$,
$a_1p_1p_2+1=a_2(2\alpha_\ell+1) +1$.  Either $2\alpha_\ell+1\vert
4m+2$ or the primitive divisor equals
$a_{12}(2\alpha_\ell+1)p_1p_2+1$.  Suppose
$$(a_{12}(2\alpha_\ell+1)+1)(x+1)={{(4k+1)^{p_1p_2}-1}\over {4k}}
\equiv p_1p_2~(mod~4) \eqno(5.12)
$$
Since $p_1,~p_2$ are odd primes, either $a_{12}\equiv 0~(mod~4)$
or $a_{12}\equiv 2~(mod~4)$, so that the primitive divisor equals
$4c(2\alpha_\ell+1)p_1p_2+1$ when $p_1p_2\equiv 1~(mod~4)$ or
$(4c+2)(2\alpha_\ell+1)p_1p_2+1$ if $p_1p_2\equiv 3~(mod~4)$.  Let
the primitive divisor $P$ have the form
$4c(2\alpha_\ell+1)p_1p_2+1$ so that
$$(4c(2\alpha_\ell+1)p_1p_2+1)(y+1)={{q_\ell^{2\alpha_\ell+1}-1}\over
 {q_\ell-1}}
\equiv 2\alpha_\ell+1~(mod~q_\ell-1) \eqno(5.13)
$$
which implies that $y+1=\kappa_2[2\alpha_\ell+\chi_2(q_\ell-1)]$
is either the multiple of an imprimitive divisor,
$\kappa_2^\prime(2\alpha_\ell+1)$ or it is the multiple of a
primitive divisor. Consider the equation
$$(4c(2\alpha_\ell+1)p_1p_2+1)\kappa_2(2\alpha_\ell+1)={{q_\ell^{2\alpha_\ell+1}-1}
                                                         \over
 {q_\ell-1}}
\eqno(5.14)
$$
with $2\alpha_\ell\vert q_\ell-1$.  If
$$(4c(2\alpha_\ell+1)p_1p_2+1)\kappa_2(2\alpha_\ell+1)\equiv
 2\alpha_\ell+1
~(mod~q_\ell-1) \eqno(5.15)
$$
$q_\ell-1\vert(4c(2\alpha_\ell+1)p_1p_2+1)\kappa_2-1$ or
$(4c(2\alpha_\ell+1)p_1p_2+1)\kappa_2-1\equiv {{q_\ell-1}\over
{2\alpha_\ell-1}}~(mod~q_\ell-1)$.   Since $2\alpha_\ell+1\vert
\kappa_2-1$, $\kappa_2=\kappa_3(2\alpha_\ell+1)+1$.  Let
$$(z(q_\ell-1)+1)(2\alpha_\ell+1)={{q_\ell^{2\alpha_\ell+1}-1}\over
 {q_\ell-1}}
\eqno(5.16)
$$
Then
$$z=2\alpha_\ell+{{q_\ell-1}\over
 {2\alpha_\ell+1}}(2\alpha_\ell-1+(2\alpha_\ell-2)
(q_\ell+1)+...+q_\ell^{2\alpha_\ell-2}+...+1) \eqno(5.17)
$$
is integer and $z(q_\ell-1)=4c(2\alpha_\ell+1)^2\kappa_3 p_1p_2
              +\kappa_3(2\alpha_\ell+1)+4c(2\alpha_\ell+1)p_1p_2$
and $2\alpha_\ell+1\vert \kappa_2-1$ so that
$\kappa_2=\kappa_3(2\alpha_\ell+1)+1$.  Let
$$(z(q_\ell-1)+1)(2\alpha_\ell+1)={{q_\ell^{2\alpha_\ell+1}-1}\over
 {q_\ell-1}}
\eqno(5.18)
$$
Then
$$z=2\alpha_\ell+{{q_\ell-1}\over
 {2\alpha_\ell+1}}(2\alpha_\ell-1+(2\alpha_\ell-2)
(q_\ell+1)+...+q_\ell^{2\alpha_\ell-2}+...+1) \eqno(5.19)
$$
is integer and $z(q_\ell-1)=4c(2\alpha_\ell+1)^2\kappa_3 p_1p_2
              +\kappa_3(2\alpha_\ell+1)+4c(2\alpha_\ell+1)p_1p_2$

In the latter case, the product of the primitive divisors takes
the form $b(2\alpha_\ell+1)+1$ so that
$\kappa_2[(2\alpha_\ell+1)+\chi_2(q_\ell-1)]=b(2\alpha_\ell+1)+1$
since $2\alpha_\ell+1$ is prime.  The two congruence relations
$$\eqalign{[(b-1)+(b(2\alpha_\ell+1)4cp_1p_2](2\alpha_\ell+1)+1&\equiv
 0
                                                       ~(mod~q_\ell-1)
\cr (b-\kappa_2)(2\alpha_\ell+1)+1&\equiv 0~(mod~q_\ell-1) \cr}
\eqno(5.20)
$$
imply
$$[\kappa_2-1+(b(2\alpha_\ell+1)+1)4cp_1p_2](2\alpha_\ell+1)\equiv 0~
(mod~q_\ell-1)
\eqno(5.21)
$$
When $2\alpha_i+1\not\vert q_\ell-1$, it follows that
$q_\ell-1\vert \kappa_2-1+(b(2\alpha_\ell+1)+1)4cp_1p_2$.  Since
every primitive divisor is congruent to 1 modulo $2\alpha_\ell+1$,
$\kappa_2= \kappa_3(2\alpha_\ell+1)+1$ and $q_\ell-1\vert
\kappa_3(2\alpha_\ell+1)+(b(2\alpha_\ell+1)+1)4cp_1p_2$.

The factorizations
$$\eqalign{
(4k+1)^{4m+2}-1&=\prod_{k=0}^{4m+1}~((4k+1)-\omega_{4m+2}^k) \cr
q_\ell^{2\alpha_\ell+1}-1
&=\prod_{k^\prime=0}^{2\alpha_i}~(q_\ell-\omega_{2\alpha_i+1}^{k^\prime})
\cr} \eqno(5.22)
$$
yield real factors which can be identified only if
$((4k+1)-1)^{t_k}=(4k)^{t_k} =(q_\ell-1)^{t_k^\prime}$ or
$(4k+2)^{t_k}=(q_\ell+1)^{t_k^\prime}$. Then,
$4k=(q_\ell-1)^{n_\ell}$, $t_k=n_\ell t_k^\prime$ or
$4k+2=(q_\ell+1)^{n_\ell}$, $t_k=n_\ell t_k^\prime$.  Since
$\sum_k~t_k=4m+2$, $\sum_k~t_k^\prime=2\alpha_i+1$,
$4m+2=n_\ell(2\alpha_\ell+1)$.  If $4m+2$ is the product of two
primes $p_1p_2$, $2\alpha_\ell+1$ must equal one of the primes,
$p_2$.

The congruence (5.15) becomes
$$[(4c(2\alpha_\ell+1)^2p_1+1)\kappa_2-1](2\alpha_\ell+1)\equiv
 0~(mod~q_\ell-1)
\eqno(5.23)
$$
If $4k=(q_\ell-1)^{p_1}$, then ${{(4k+1)^{p_1p_2}-1}\over {4k}}=
{{((q_\ell-1)^{p_1}+1)^{p_1p_2}-1}\over {(q_\ell-1)^{p_1}}}
=p_1p_2+\left({{p_1p_2}\atop 2}\right)(q_\ell-1)
+...+p_1p_2(q_\ell-1)^{p_1(p_1p_2-2)}+(q_\ell-1)^{p_1(p_1p_2-1)} $
When $2\alpha_i+1\vert q_\ell-1$, $2\alpha_i+1\vert
p_1p_2+\left({{p_1p_2}\atop 2}\right)(q_\ell-1)
+...+p_1p_2(q_\ell-1)^{p_1(p_1p_2-2)}+(q_\ell-1)^{p_1(p_1p_2-1)}$.
However, ${{q_\ell^{2\alpha_\ell+1}-1}\over {q_\ell-1}}\equiv
2\alpha_\ell+1~ (mod~q_\ell-1)$, whereas ${{(4k+1)^{p_1p_2}-1}\over
{4k}} \equiv p_1(2\alpha_\ell+1)~(mod~q_\ell-1)$, $p_1>1$.  The
repunits have the same prime divisors only if $p_1\equiv
1~(mod~q_\ell-1)$, which implies that $p_1=\rho (q_\ell-1)+1$.  Then
${{((q_\ell-1)^{\rho(q_\ell-1)}
+1)^{\rho(q_\ell-1)(2\alpha_\ell+1)-1}}\over
{(q_\ell-1)^{\rho(q_\ell-1)}}}$ includes at least seven new factors.
If $4k+2=(q_\ell+1)^{p_1}$, then ${{(4k+1)^{p_1p_2}-1}\over
{4k}}\equiv {{(2^{p_1}-1)^{p_1p_2}-1}\over
{2^{p_1}-2}}~(mod~q_\ell-1)$.  If $2^{p_1}-1\equiv
1~(mod~q_\ell-1)$, $1+(2^{p_1}-1)+...+(2^{p_1}-1)^{p_1p_2-1}\equiv
p_1p_2\equiv 2\alpha_\ell+1~ (mod~q_\ell-1)$ only if $p_1\equiv
1~(mod~q_\ell-1)$.  When $2^{p_1}-1\equiv 2n+1~(mod~q_\ell-1)$,
${{(2^{p_1}-1)^{p_1p_2}-1}\over {2^{p_1}-2}} \equiv
{{(2n+1)^{p_1p_2}-1}\over {2n}}$ which is not divisible by
$2\alpha_i+1$ if $p_2 \not 2n$.  When $2\alpha_i+1\vert n$,
${{(2n)^{p_1p_2}-1}\over {2n}}\equiv p_1p_2~(mod~2n)$ which would be
congruent to $2\alpha_i+1$ only if $p_1\equiv 1~(mod~2n)$ so that
$p_1\ge 4\alpha_i+3$. The exponent is a product of a minimimum of
three primes and introduces at least seven different divisors in the
product of repunits.  A similar conclusion is obtained if
$4m+2=p_1..p_s,~s\ge 3$.

If every prime divisor of ${{(4k+1)^{4m+2}-1}\over {4k}}$ is
distinct from the factors of
$\prod_{i=1}^\ell~{{q_i^{2\alpha_i+1}-1}\over {q_i-1}}$, the
following equations are obtained
$$\eqalign{{{q_i^{2\alpha_i+1}-1}\over {q_i-1}}&=(4k+1)^{h_i}
                                        q_{j_i}^{2\alpha_{j_i}}
\cr {{(4k+1)^{4m+2}-1}\over {4k}}&=2 \cr \sum_{i=1}^\ell~h_i&=4m+1
\cr} \eqno(5.24)
$$
which only has the solution $k=0$.

The number of equal prime divisors in
${{q_{i_0}^{2\alpha_{i_0}+1}-1}\over {q_{i_0}-1}}$ and
${{(4k+1)^{4m+2}-1}\over {4k}}$ can be chosen to be greater than 1,
but equality of $\sigma(N)$ and $2N$ implies that each distinct
prime divisor $q_{j_i}$ of the repunit ${{q_i^{2\alpha_i+1}-1}\over
{q_i-1}}$ appears in the factorization with the exponent
$2\alpha_{j_i}$.  Consequently, it would be inconsistent with the
perfect number condition for divisors of
${{q_{i_0}^{2\alpha_{i_0}+1}-1}\over {q_{i_0}-1}}$ other than
$q_{j_{i_0}}$ to exist.

If the prime divisor $q_{j_{i_0}}$ of
${{q_{i_0}^{2\alpha_{i_0}+1}-1}\over {q_{i_0}-1}}$ is a factor of
${{(4k+1)^{4m+2}-1}\over {4k}}$, then one formulation of the
perfect number condition for the integer
$N=(4k+1)^{4m+1}q_i^{2\alpha_i}$ is
$$\eqalign{{{q_i^{2\alpha_i+1}-1}\over {q_i-1}}&=(4k+1)^{h_i}
q_{j_i}^{2\alpha_{j_i}}~~~~~~~~i\ne i_0
\cr {{q_{i_0}^{2\alpha_{i_0}+1}-1}\over {q_{i_0}-1}}&=
q_{j_{i_0}}^{h_{j_{i_0}}^\prime}
\cr {{(4k+1)^{4m+2}-1}\over
{4k}}&=2q_{i_0}^{2\alpha_{j_{i_0}}-h_{j_{i_0}}^\prime} \cr}
\eqno(5.25)
$$
The last relation in equation (5.25) has no solutions with $k\ge
1$ since ${{(4k+1)^{4m+2}-1}\over {4k}}$ has at least four
different prime divisors, $2$,$2k+1$, the prime factors of
${{(4k+1)^{2m+1}-1}\over {4k}}$ and its primitive divisors.
Furthermore, as the number of prime divisors of
${{(4k+1)^{4m+2}-1}\over {4k}}$ is less than seven,
${{q_{i_0}^{2\alpha_{i_0}+1}-1}\over {q_{i_0}-1}}$ also should
have a different prime factor which is contrary to the equation
(5.25).

Since $k\ge 1$ in the decomposition
$N=(4k+1)^{4m+1}\prod_{i=1}^\ell~ q_i^{2\alpha_i}$, the quotient
\hfil\break ${{(4k+1)^{2m+1}-1}\over {4k}},~k\ge 1$ has a distinct
prime divisor from the factors of
$\prod_{i=1}^\ell~{{q_i^{2\alpha_i+1}-1}\over {q_i-1}}$ because of
its existence in the factorization of ${{(4k+1)^{p_1(2m+1)}-1}\over
{4k}}$ by Theorem 1.  There would be then at least $\ell+3$ prime
factors of $\sigma(N)$ implying that $N$ is not an odd perfect
number.

\line{\hfil{\qed}} \vfill\eject \noindent {\bf 6. A Proof of the Odd
Perfect Number Conjecture} \vskip 10pt \noindent {\bf Theorem 3.}
There are no odd perfect numbers. \vskip 5pt \noindent {\bf Proof.}
The exponents $2\alpha_i+1$, $i=1,...,\ell$ will be assumed to be
prime, since repunits with composite exponents have at least three
prime divisors [11].  Additionally, it has been established that
there are at most two solutions for fixed $x$ and $y$ if $y\ge 7$ [6].

If $a={{y-1}\over \delta}$, $b={{x-1}\over \delta}$, $c={{y-x}\over
 \delta}$,
$\delta=gcd(x-1,y-1)$ and $s$ is the least integer such that
$x^s\equiv 1~(mod~by^{n_1})$, where $(x,y,m_1,n_1)$ is a presumed
solution of the Diophantine equation, then $m_2-m_1$ is a multiple
of $s$ if $(x,y,m_2,n_2)$ is another solution with the same bases
$x,y$ [6].  As $x < y$, $x$ can be identified with $q_j$, $y$ with
with $q_i$ and $\delta$ with $gcd(q_j-1,q_i-1)$. Given that the
first exponents are $2\alpha_i+1$ and $2\alpha_j+1$, the constraint
on a second exponent of $q_j$ then would be $m_2-(2\alpha_j+1)=\iota
\varphi\left(\left({{q_j-1}\over \delta}\right)
q_i^{2\alpha_i+1}\right)$ where $\iota\in {\bf Q}$, with $\iota\in
{\bf Z}$ when $q_j\not\vert q_i-1$.  This constraint can be extended
to $m_1=n_1=1$, since it follows from the equations
$(y-1)x^{m_i}-(x-1)y^{n_i}=y-x$, $ax^{m_i}-by^{n_i}=c$, which hold
also for these values of the exponents. Based on the trivial
solution to these equations for an arbitrary pair of odd primes
$q_i,q_j$, the first non-trivial solution for the exponent of $q_j$
would be greater than or equal to than $1+\iota (q_i-1)\varphi
\left({{q_j-1}\over \delta}\right)$. This exponent introduces new
prime divisors which are factors of ${{q_j^{1+ \iota
(q_i-1)\varphi\left(\left({{q_j-1}\over \delta}\right)
\right)}-1}\over {q_j-1}}$ and therefore does not minimize the
number of prime factors in the product of repunits.

Imprimitive prime divisors can be introduced into equations
generically relating the two unequal repunits and minimizing the
number of prime factors in the product \hfil\break
${{q_i^{2\alpha_i+1}-1}\over {q_i-1}}{{q_j^{2\alpha_j+1}-1}\over
{q_j-1}}$. Given an odd integer
$N=(4k+1)^{4m+1}\prod_{i=1}^\ell~q_i^{2\alpha_i}$ the least number
of unmatched prime divisors in $\sigma(N)$ will be attained if there
are pairs $(q_i,q_j)$ satisfying one of the three relations
$$\eqalign{{{q_i^{2\alpha_i+1}-1}\over {q_i-1}}&=(2\alpha_i+1)\cdot
{{q_j^{2\alpha_j+1}-1}\over {q_j-1}} \cr (2\alpha_j+1)\cdot
{{q_i^{2\alpha_i+1}-1}\over {q_i-1}}&={{q_j^{2\alpha_j+1}-1}\over
{q_j-1}} \cr (2\alpha_j+1)\cdot {{q_i^{2\alpha_i+1}-1}\over
{q_i-1}}&=(2\alpha_i+1) \cdot {{q_j^{2\alpha_j+1}-1}\over {q_j-1}}
\cr} \eqno(6.1)
$$

As ${{q_j^{2\alpha_j+1}-1}\over {q_j-1}}$ would not introduce any
additional prime divisors if the first relation in equation (6.1)
is satisfied, the product of two pairs of repunits of this kind,
with prime bases $(q_i,q_j)$, $(q_k,q_{k^\prime})$, yields a
minimum of three distinct prime factors, if two of the repunits
are primes. In the second relation, $(q_i,2\alpha_i+1)$ and
$(q_j,2\alpha_j+1)$ are interchanged, whereas in the the third
relation, an additional prime divisor is introduced when
$2\alpha_i+1\ne 2\alpha_j+1$.  Equality of $\sigma(N)$ and 2N is
possible only when the prime divisors of the product of repunits
also arise in the decomposition of N. If the two repunits
${{q_j^{2\alpha_j+1}-1}\over {q_j-1}}$ and
${{q_{k^\prime}^{2\alpha_{k^\prime}+1}-1}\over {q_{k^\prime}-1}}$
are prime, they will be bases for new repunits
$$\eqalign{
{{\left[{{q_j^{2\alpha_j+1}-1}\over
{q_j-1}}\right]^{2\alpha_{n_1}+1}-1} \over
{{{q_j^{2\alpha_j+1}-1}\over {q_j-1}}-1}}={{q_j-1}\over
{q_j(q_j^{2\alpha_j}-1)}} &\cdot
\left[\left[{{q_j^{2\alpha_j+1}-1}\over
{q_j-1}}\right]^{2\alpha_{n_1}+1}-1 \right] \cr
{{\left[{{q_{k^\prime}^{2\alpha_{k^\prime}+1}-1}\over
{q_{k^\prime}-1}} \right]^{2\alpha_{n_2}+1}-1} \over
{{{q_{k^\prime}^{2\alpha_{k^\prime}+1}-1}\over
{q_{k^\prime}-1}}-1}} ={{q_{k^\prime}-1}\over
{q_{k^\prime}(q_{k^\prime}^{2\alpha_{k^\prime}}-1)}} &\cdot
\left[\left[{{q_{k^\prime}^{2\alpha_{k^\prime}+1}-1}\over
{q_{k^\prime}-1}} \right]^{2\alpha_{n_2}+1}-1 \right] \cr}
\eqno(6.2)
$$
These new quotients are either additional prime divisors or
satisfy relations of the form
$$
{{\left[{{q_j^{2\alpha_j+1}-1}\over
{q_j-1}}\right]^{2\alpha_{n_1}+1}-1} \over
{{{q_j^{2\alpha_j+1}-1}\over
{q_j-1}}-1}}=(2\alpha_{n_1}+1){{q_{t_1}^{2\alpha_{t_1} +1}-1}\over
{q_{t_1}-1}} \eqno(6.3)
$$
to minimize the introduction of new primes.  Furthermore,
$2\alpha_{n_1}+1$ is the factor of lesser magnitude,
$${{q_t^{2\alpha_t+1}-1}\over {q_t-1}} > \Biggl[
{{q_j-1}\over {q_j(q_j^{2\alpha_j}-1)}} \cdot
\left[\left[{{q_j^{2\alpha_j+1}-1}\over
{q_j-1}}\right]^{2\alpha_{n_1}+1}-1 \right]\Biggr]^{1\over 2}
\eqno(6.4)
$$
The number of prime divisors of N is minimized if the repunit
${{q_t^{2\alpha_t+1}-1} \over {q_t-1}}$ is a prime.  However, it
is then the basis of another repunit
$${{\left[{{q_t^{2\alpha_t+1}-1}\over
 {q_t-1}}\right]^{2\alpha_{n_3}+1}-1}
\over {\left[{{q_t^{2\alpha_t+1}-1}\over {q_t-1}}\right]-1}}
> {{\left[{{q_j^{2\alpha_j+1}-1}\over
 {q_j-1}}\right]^{2\alpha_{n_1}+1}-1}
\over {{{q_j^{2\alpha_j+1}-1}\over {q_j-1}}-1}} \eqno(6.5)
$$
since $\alpha_{n_3}\ge 1$.  An infinite sequence of repunits of
increasing magnitude is generated, implying the non-existence of
odd integers $N$ with prime factors satisfying one of the
relations in equation (6.1).

If the third relation in equation (6.1) holds,
$$(2\alpha_t+1){{\left[{{q_j^{2\alpha_j+1}-1}\over
 {q_j-1}}\right]^{2\alpha_{n_1}+1}-1}
\over {{{q_j^{2\alpha_j+1}-1}\over {q_j-1}}-1}}=(2\alpha_{n_1}+1)
{{q_{t_1}^{2\alpha_{t_1} +1}-1}\over {q_{t_1}-1}} \eqno(6.6)
$$
The inequality (6.4) is still satisfied, and again, repunits of
increasing magnitude are introduced in the sum of divisors.

If the repunits ${{q_j^{2\alpha_j+1}-1}\over {q_j-1}}$ and
${{q_{\ell^\prime}^{2\alpha_{\ell^\prime}+1}-1}\over
{q_{\ell^\prime}-1}}$ are not prime, then the product of the four
repunits with bases $q_i,q_j,q_k,q_{k^\prime}$ would have a minimum
of five different prime divisors.  The product of
${{(4k+1)^{4m+2}-1}\over {4k}}$, with at least two distinct prime
divisors, and $\prod_{i=1}^\ell~{{q_i^{2\alpha_i+1}-1}\over
{q_i-1}}$ possesses a minimum of $\ell+3$ different prime factors
and the perfect number condition cannot be satisfied.

When none of the three relations hold in equation (6.1) hold, by
Theorem 1, there again exists a primitive divisor which is not a
common divisor of both repunits. It can be assumed that the last
repunit ${{q_\ell^{2\alpha_\ell+1}-1}\over {q_\ell-1}}$ has a prime
exponent since it will introduce at least three prime divisors if
the exponent is composite. The new prime divisor may be denoted
$q_{j_\ell}$ if it does not equal $4k+1$, and interchanging $q_\ell$
with $q_{\bar i}$ and $4k+1$, it can be deduced that
$U_{4m+2}(4k+2,4k+1)\prod_{{i=1}\atop {i\ne {\bar i}}}^\ell
U_{2\alpha_i+1} (q_i+1,q_i)$ will not be divisible by $j_{\bar i}\in
1,...,\ell$, $j_{\bar i} \ne {\bar i}$, with the exception of one
value $i_0$.  The prime $q_{j_0}$ will not be a factor of
$\prod_{i=1}^\ell~U_{2\alpha_i+1}(q_i+1,q_i)$.  Then
$$\eqalign{{{q_{\bar i}^{2\alpha_{\bar i}+1}-1}\over {q_{\bar i}-1}}&=
q_{j_{\bar i}}^{h_{j_{bar i}}}~~~~~~~~~{\bar i}\ne i_0 \cr
{{q_{i_0}^{2\alpha_{i_0}+1}-1}\over {q_{i_0}-1}}&=(4k+1)^{4m+1}
\cr {{(4k+1)^{4m+2}-1}\over {4k}}&=2q_{j_{i_0}}^{h_{j_{i_0}}} \cr}
\eqno(6.7)
$$
The last equation is equivalent to
$$\eqalign{{{(4k+1)^{2m+1}-1}\over {4k}}&=y_1^2
\cr {{(4k+1)^{2m+1}+1}\over 2}&=y_2^2 \cr
y_1y_2&=q_{j_{i_0}}^{{h_{j_{i_0}}}\over 2} \cr} \eqno(6.8)
$$
which has no integer solutions for $k$ and $m$.  There are no
prime sets $\{4k+1;q_1,...,q_\ell\}$ which satisfy these
conditions with $h_{j_{\bar i}}$ and $h_{j_{i_0}}$ even.  Thus,
there are no odd integers
$N=(4k+1)^{4m+1}\prod_{i=1}^\ell~q_i^{2\alpha_i}$ such that the
primes and exponents $\{q_i,2\alpha_i+1\}$ satisfy none of the
relations in equation (6.1) and equality between $\sigma(N)$ and
$2N$.  For all odd integers, ${{\sigma(N)}\over N}\ne 2$.
\hfil\break \line{\hfil{\qed}}

\vfill\eject {\bf References}
\vskip 10pt
\item{[1]} A. S. Bang,
{\it Taltheoretiske Undersogelser}, Tidsskrift for Mathematik 5
(1886), 265-284.
\item{[2]} Yu. Bilu, G. Hanrot and P. M. Voutier,
{\it Existence~of~ primitive~divisors~of} \hfil\break {\it
Lucas~and~Lehmer~numbers}, J. f{\"u}r die reine und angewandte
Mathematik \hfil\break 539 (2001), 75-122.
\item{[3]} F. de Bessy (1657), in:
Oeuvres de Huygens, II, Correspondence, No. 389; Ouvres de Fermat, 3,
Gauthier-Villars, Paris, 1896, p.567.
\item{[4]} G. D. Birkhoff
and H. S. Vandiver, {\it On~the~integral~of ~$a^n-b^n$}, Ann. Math.
5 no. 2 (1904), 173-180.
\item{[5]} R. P. Brent, G. L. Cohen and H. J. J. te Riele,
{\it Improved~Techniques~for~Lower~Bounds}
\hfil\break
{\it for~Odd~Perfect~Numbers},
Math. Comp.  57 (1991) 857-868.
\item{[6]} Y. Bugeaud and T. N. Shore, {\it On~the~Diophantine~Equation
~${{x^m-1}\over {x-1}}={{y^n-1}\over {y-1}}$}, Pac. J. Math. 207(1)
(2002) 61-75.
\item{[7]} E. Catalan, Mathesis 8 (1888) 112-113; Mem. Soc. Sc.
Li{\`e}ge (2) 15 (1888) 205-207.
\item{[8]} R. D. Carmichael, {\it
Multiply~Perfect~Odd~numbers~with~Three~Prime~Factors}, Amer. Math.
Monthly 13 (1906), 35-36.
\item{[9]} R. D. Carmichael, {\it
Multiply~Perfect~Numbers~of~Four~Different~Primes}, Ann. Math. 8
(1907), 149-158.
\item{[10]} G. L. Cohen, {\it
 On~the~Largest~Component~of~an~Odd~Perfect
~Number},
J. Austral. Math. Soc. 42 (1987) 280-286.
\item{[11]} S. Davis, {\it
A~Rationality~Condition~for~the~Existence~of~ Odd~Perfect~Numbers},
Interna- \hfil\break tional Journal of Mathematics and Mathematical
Sciences, Vol. 2003, No. 20 (2003), 1261-1293.
\item{[12]} R. Descartes to M. Mersenne (1638), Oeuvres de
Descartes, II, eds. C. Adam and P. Tannery, Cerf, Paris, 1898,
p. 429.
\item{[13]} L. Euler,
{\it Tractatus~de~Numerorum~Doctrina}, $\S 109$ in: Opera Omnia I 5
Auctoritate et Impensis Societatis Scientarum Naturalium Helveticae,
Genevae, MCMXLIV.
\item{[14]} R. Goormaghtigh, {\it
Nombres~tel~que~$A=1+x+x^2+...+x^m=1+y+y^2+ ...+y^n$}, \hfil\break
L'Intermediare des Mathematiciens
 3 (1892), 265-284.
\item{[15]} T. Goto and Y. Ohno, {\it Odd~Perfect~Numbers~have
~a~Prime~Factor~Exceeding~$10^8$}, preprint (2006).
\item{[16]} O. Grun, {\it {\"U}ber~ungerade~Volkommene~Zahlen}, Math.
Zeit. {\bf 55} (1952) 353-354.
\item{[17]} P. Hagis Jr., {\it An~Outline~of~a~Proof~that~Every~Odd
~Perfect~Number~has~at~Least~Eight}
\hfil\break
{\it Prime~Factors}, Math. Comp.
{\bf 34} (1983) 1027-1032.
\item{[18]} P. Hagis Jr., {\it Sketch~of~a~Proof~that~an~Odd~Perfect
~Number~Relatively~Prime~to~3~has~at}
\hfil\break
{\it Least~Eleven~Prime~Factors},
Math. Comp. {\bf 40} (1983) 399-404.
\item{[19]}  K. G. Hare, {\it
 More~on~the~Total~Number~of~Prime~Factors~of~
an~Odd~Perfect~Number}, Math. Comp. {\bf 74} (2005) 1003-1008.
\item{[20]} K. Hare, {\it
 New~Techniques~for~Bounds~on~the~Total~Number~of
~Prime~Factors~of~an~Odd}
\hfil\break
{\it Perfect~Number}, math.NT/0501070.
\item{[21]} D. E. Iannucci, {\it The~Second~Largest~Prime~Divisors~of
~an~Odd~Perfect~Number~exceeds~Ten~Thousand}, Math. Comp. {\bf 68}
 (1999)
 1749-1760.
\item{[22]} D. E. Iannnucci, {\it The~Third~Largest~Prime~Divisor~of
~an~Odd~Perfect~Number~exceeds~One~Hundred}, Math. Comp. {\bf 69}
 (2000)
 867-879.
\item{[23]} M. Jenkins, {\it
 Odd~Perfect~Numbers~have~a~Prime~Factor~Exceeding~$10^7$}, Math.
Comp. {\bf 72} (2003) 1549-1554.000
\item{[24]} M. Kishore, {\it
 Odd~Integers~$N$~with~Five~Distinct~Prime~Factors
~for~which~$2-10^{-12}<{{\sigma(N)}\over N}<2+10^{-12}$}, Math. Comp.
 {\bf 32}
(1978) 303-309.
\item{[25]} M. Kishore, {\it Odd~Perfect~Numbers~not~Divisible~by~$3$},
 Math. Comp.
{\bf 40} (1983) 405-411.
\item{[26]} N. Koblitz, {\it
 p-adic~Numbers,~p-adic~Analysis~and~Zeta~Functions}
(springer-Verlag, 1977)
\item{[27]} P. P. Nielsen, {\it Odd~Perfect~Numbers~have~at~Least
~Nine~ Different~Prime~Factors}, Math. Comp. {\bf 76} (2007) 2109-2126.
\item{[28]} K. K. Norton, {\it
 Remarks~on~the~Number~of~Factors~of~an~Odd~Perfect~Number},
Acta Arith. {\bf 6} (1960) 365-374.
\item{[29]} Y. Nesterenko and T. N. Shorey, {\it
 On~the~Diophantine~Equation
~${{x^m-1}\over {x-1}}={{y^n-1}\over {y-1}}$}, Acta Arith. {\bf 83}
 (1998) 381-399.
\item{[30]} O. Ore, {\it On~the~Averages~of~the~Divisors~of~a~Number},
 Amer.
Mat. Monthly, 55 (1948) 615-619.
\item{[31]} T. N. Shorey. {\it On~the~Equation~$ax^m-by^n=k$}, Indag.
 Math. {\bf 48} (1986)
353-358.
\item{[32]}  C. L. Stewart, {\it
 On~Divisors~of~Fermat,~Fibonacci,~Lucas~and~Lehmer~Numbers},
Proc. London Math. Soc. Series 3 {\bf 35} (1977) 425-477.
\item{[33]} J. Sylvester, {\it Sur~les~Nombres~Parfaits}, C. R. CVI
 (1888) 403-405.
\item{[34]} T. Yamada, {\it Odd~Perfect~Numbers~have~a~Special~Form},
 Colloq. Math.
{\bf 103} (2005) 303-307.
\item{[35]} K. Zsigmondy, {\it Zur~Theorie~der~Potenzreste}, Monatsh.
 Math. {\bf 3} (1892)
265-284.

\end